\RequirePackage{fix-cm}
\documentclass[smallextended]{svjour3}       
\smartqed  
\usepackage{graphicx}

\usepackage{enumitem}
\usepackage{csquotes}
\usepackage{amsmath,amssymb,amsfonts}
\usepackage{latexsym}
\usepackage{algorithmic}
\usepackage{textcomp}
\usepackage[table]{xcolor}
\usepackage{xcolor}
\def\BibTeX{{\rm B\kern-.05em{\sc i\kern-.025em b}\kern-.08em
    T\kern-.1667em\lower.7ex\hbox{E}\kern-.125emX}}
    
\usepackage{booktabs,makecell,siunitx,eqparbox}

\usepackage{hyperref}
\usepackage{listings}
\usepackage{adjustbox}
\usepackage{multirow}
\usepackage{rotating}
\usepackage{tcolorbox}
\usepackage{subcaption}
\usepackage{adjustbox}

\definecolor{codegreen}{rgb}{0,0.6,0}
\definecolor{codegray}{rgb}{0.5,0.5,0.5}
\definecolor{codepurple}{rgb}{0.58,0,0.82}
\definecolor{backcolour}{rgb}{0.95,0.95,0.92}

\lstdefinestyle{mystyle}{
    backgroundcolor=\color{backcolour},   
    commentstyle=\color{codegreen},
    keywordstyle=\color{blue},
    numberstyle=\tiny\color{codegray},
    stringstyle=\color{codepurple},
    basicstyle=\ttfamily\footnotesize,
    breakatwhitespace=false,         
    breaklines=true,                 
    captionpos=b,                    
    keepspaces=true,                 
    numbers=left,                    
    numbersep=5pt,                  
    showspaces=false,                
    showstringspaces=false,
    showtabs=false,                  
    tabsize=2
}

\usepackage[numbers]{natbib}
\usepackage[normalem]{ulem}

\begin{document}

\title{Common Challenges of Deep Reinforcement Learning Applications Development: An Empirical Study \thanks{This work was supported by: Fonds de Recherche du Québec (FRQ), the Canadian Institute for Advanced Research (CIFAR) as well as the DEEL project CRDPJ 537462-18 funded by the National Science and Engineering Research Council of Canada (NSERC) and the Consortium for Research and Innovation in Aerospace in Québec (CRIAQ), together with its industrial partners Thales Canada inc, Bell Textron Canada Limited, CAE inc and Bombardier inc.}
}

\titlerunning{Common Challenges of DRL Applications Development}        

\author{Mohammad Mehdi Morovati \and
        Florian Tambon \and
        Mina Taraghi \and
        Amin Nikanjam \and 
        Foutse Khomh
}


\institute{Mohammad Mehdi Morovati \and
            Florian Tambon \and
            Mina Taraghi \and
           Amin Nikanjam
              \and
            Foutse Khomh \at
            SWAT Lab., Polytechnique Montréal, Montréal, Canada\\
            \email{\{mehdi.morovati,florian-2.tambon,mina.taraghi,amin.nikanjam,foutse.khomh\}@polymtl.ca}\\
}

\date{Received: date / Accepted: date}

\maketitle

\begin{abstract}
Machine Learning (ML) is increasingly being adopted in different industries. 
Deep Reinforcement Learning (DRL) is a subdomain of ML used to produce intelligent agents. Despite recent developments in DRL technology, 
the main challenges that developers face in the development of DRL applications are still unknown. To fill this gap, in this paper, we conduct a large-scale empirical study of \textbf{927} DRL-related posts extracted from Stack Overflow, the most popular Q\&A platform in the software community. 
Through the process of labeling and categorizing extracted posts, we created a taxonomy of common challenges encountered in the development of DRL applications,
along with their corresponding popularity levels. 
This taxonomy has been validated through a survey involving 65 DRL developers. Results show that at least $45\%$ of developers experienced 18 of the 21 challenges identified in the taxonomy. 
The most frequent source of difficulty during the development of DRL applications are \textit{Comprehension}, \textit{API usage}, and \textit{Design problems}, while \textit{Parallel processing}, and \textit{DRL libraries/frameworks} are classified as the most difficult challenges to address, with respect to the time required to receive an accepted answer. 
We hope that the research community will leverage this taxonomy to develop efficient strategies to address the identified challenges and improve 
the quality of DRL applications. 

\end{abstract}

\section{Introduction}
Reinforcement Learning (RL) has begun making its mark across a range of industrial sectors, from autonomous vehicles~\cite{aradi2020survey} and traffic engineering~\cite{xiao2021leveraging} to healthcare systems~\cite{yu2021reinforcement}. Recently we have been also witnessing an increasing adoption of 
RL to solve different software engineering tasks, from automatic code improvement~\cite{wan2018improving}, to test case prioritization~\cite{bagherzadeh2021reinforcement}, and program debloating~\cite{heo2018effective}. 
Reinforcement Learning differs significantly from other subcategories of Machine Learning (ML) such as supervised and unsupervised learning, as it includes an agent that interacts with an environment to learn how to perform a sequence of actions leading to the best cumulative final rewards~\cite{nikanjam2022faults}. In other words, in RL, an agent learns to act in a way that modifies its behavior gradually to achieve the best final result; which makes traditional software quality assurance techniques inadequate for RL. 


Deep Reinforcement Learning (DRL) is an integration of Deep Learning (DL) and RL, also known as Deep RL, to address challenges, such as 
high-dimension input data~\cite{arulkumaran2017deep}. 
Combining DL and RL enables DRL to discover compact low-dimensional representations of high-dimensional data automatically~\cite{arulkumaran2017deep}. 


Although there exist studies on testing and debugging RL programs~\cite{zolfagharian2022search,tambon2023mutation}, the main challenges and obstacles that developers face while developing RL applications are still unclear~\cite{zhang2020deep}. 
Moreover, because of basic differences between the paradigm of traditional software applications and ML applications~\cite{morovati2023bugs,islam2020repairing}, it is expected that developers of ML applications face different types of challenges in the implementation process of such applications. 
Thus, DRL developers may face different challenges from other types of software systems (including traditional software systems as well as other subcategories of ML applications)
\cite{nguyen2020deep,du2021survey,dulac2021challenges}.

As an example, Listing.\ref{fig:so_sample_drl} shows a SO post (\href{https://stackoverflow.com/questions/70562317}{70562317}) related to a DRL application, representing a challenge in implementing the method to choose an optimal action which is specific to DRL development and differs from ML and DL development challenges.


    \begin{lstlisting}[language=Python, float, floatplacement=H, label=fig:so_sample_drl,caption=SO post (\href{https://stackoverflow.com/questions/70562317}{70562317}) showing a challenge in the development of the RL action.]
def act(self, some_input, state):
    mu, var, state_value = self.model(some_input, state)
    # mu contains info required for gradient
    mu = mu.data.cpu().numpy()
    # mu is detached and now has forgot all the operations performed in "self.action_head"
    sigma = torch.sqrt(var).data.cpu().numpy()
    action = np.random.normal(mu, sigma)
    action = np.clip(action, 0, 1)
    action = torch.from_numpy(action/1000)
    return action, state_value\end{lstlisting}

Although there exist some studies regarding challenges in the development of DL~\cite{zhang2019empirical,rao2018deep}, ML applications~\cite{lwakatare2019taxonomy,de2019understanding},
to the best of our knowledge there is no study on the challenges that developers face when developing DRL applications.  
The study by Yahmed et al.~\cite{yahmed2023deploying} is the most closely related work to this research. It examines the challenges that developers face during the deployment process of DRL systems but does not consider the challenges occurring in the early development phases prior to deployment. 
In this study, we examine the following 
research questions:\\
\textit{\textbf{RQ1.} What are the common challenges of DRL application development?}\\
\textit{\textbf{RQ2.} How are the identified challenges perceived by DRL practitioners?}\\
\textit{\textbf{RQ3.} Are DRL application development challenges language- and/or framework-specific?}\\
To answer these research questions, we manually examined and categorized 927 Stack Overflow (SO) posts that are related to DRL development. We report our results as a taxonomy of challenges in DRL application development. Besides, we conducted a survey of DRL developers/practitioners to validate our findings. 
Moreover, we investigated the 
dependency of the identified challenges on 
programming languages and libraries/frameworks used for DRL development.
The contributions of this study are summarized as follows. 
\begin{itemize}
    \item We provide the first large-scale empirical study of the challenges in the development of DRL applications,
    \item We categorize challenges in DRL application development and propose a taxonomy, 
    \item We conduct a survey with DRL practitioners to validate the identified common challenges of DRL application development, 
    \item We examine the relationship between the identified challenges and the programming languages and libraries/frameworks used to develop DRL applications.
\end{itemize}


\textbf{The rest of the paper is as follows.} We describe the methodology of our study in Section \ref{sec:methodology}. 
In Section \ref{sec:results}, we report our findings including the taxonomy of DRL development challenges. 
Section \ref{sec:discussion} discusses the implications of the highlighted findings. 
Afterward, we review related works in Section \ref{sec:related_work}. Threats to the validity of our research, and conclusion/future works are discussed in Section \ref{sec:validity} and Section \ref{sec:conclusion}, respectively. 

\section{Methodology}
\label{sec:methodology}
\begin{figure*}[t]
    \centering
    \includegraphics[width=0.8\textwidth]{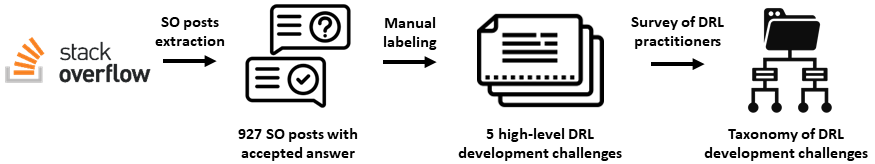}
    \caption{High-level view of the used methodology.}
    \label{fig:method}
    \vspace{-1em}
\end{figure*}

This section describes the methodology we follow in this study. This methodology is illustrated in Fig.\ref{fig:method}. 

\subsection{Extracting Posts from Stack Overflow (SO)}
\label{subsec:so_extract}
We rely on Stack Overflow (SO) as the main source of information in this study; similar to several previous studies which utilized data exclusively obtained from SO for analyses~\cite{zhang2019empirical,alshangiti2019developing,hamidi2021towards}.
SO is known as the largest technical question and answer (Q\&A) website creating a public knowledge base in various areas~\cite{zhu2022empirical}, with 23.4 million questions and 19.6 million users as of December 2022~\cite{so_data_dump}. In the software development community, SO provides a platform for developers to exchange about coding issues; improving their coding knowledge. To extract SO posts related to DRL, we use Stack Exchange Data Explorer\footnote{https://data.stackexchange.com/stackoverflow/query} which provides access to up-to-date SO data. 
Overall, we use a list of DRL-related tags and keywords to collect DRL-related SO posts. Listing.\ref{fig:so_query} presents an example of a query used to collect SO posts containing \textit{`deep-learning'} and \textit{`reinforcement-learning'} tags at the same time. 

    \begin{lstlisting}[language=SQL, label=fig:so_query,caption=Sample query to extract SO posts that have \textit{`deep-learning'} and \textit{`reinforcement-learning'} tags simultaneously.]
SELECT * FROM Posts
WHERE Tags LIKE '%<deep-learning>%' 
AND Tags LIKE '%<reinforcement-learning>%'
\end{lstlisting}

To gather the list of DRL-related tags and keywords, we follow previous study~\cite{alshangiti2019developing} using a snowballing approach in which we start with posts that have \textit{`deep-learning'} and \textit{`reinforcement-learning'} tags, simultaneously. 
In the next step, we collect all tags assigned to the SO posts gathered in the previous step. Then, we include DRL-related tags to our list of tags (e.g., \textit{`dqn'}). We continue this process and expand the list of DRL-related tags until we are unable to add any new tags to our list. Besides, we create a list of DRL-related keywords based on the list of collected DRL-related tags. Firstly, we include all DRL-related tags (such as `reinforcement learning') to the list of DRL-related keywords. Moreover, we add expanded forms of DRL-related tags which are acronyms. As an example, we add \textit{`Deep Q-Learning'} which is the expanded form of \textit{`dqn'}.
The complete list of tags and keywords used to extract SO posts are available in our replication package~\cite{replication_package}. In summary, we collected SO posts that meet at least one of the following criteria:

\begin{itemize}
    \item Posts having one of the identified tags (e.g., `\textit{drl}', `\textit{dqn}', etc.)
    \item Posts with a combination of identified tags (e.g., combination of `\textit{deep-learning}' and `\textit{reinforcement-learning}')
    \item Posts with a combination of identified tags and keywords in their title or body (e.g., `\textit{reinforcement-learning}' tag and `\textit{deep}' in the post title)
    \item Posts including identified keywords in their title or body (e.g., `\textit{drl}')
\end{itemize}

After extracting all posts and removing duplicates, we obtained 3,083 posts. Then, we filtered out the posts without an accepted answer which leaves us with 927 posts. We chose to remove posts without an accepted answer, similar to previous studies~\cite{nikanjam2022faults,he2023representation} because the correctness of any of such responses would not be inferable, which could potentially bias our results.
We also collected information about the time taken by each post to receive an accepted answer and used it as an indicator of the level of difficulty of the question, similar to the approach employed in previous 
works~\cite{haque2020challenges,zahedi2020mining}.

\subsection{Manual Inspection}
During this step, a team of four raters (three Ph.D. candidates and a senior research staff who all are practitioners of DRL development) is responsible for labeling the collected SO posts. Following a methodology similar to prior works~\cite{humbatova2020taxonomy,islam2019comprehensive}, we split the collected SO posts into 10 parts, each of which is inspected in a dedicated labeling round. All the discussions and referenced source codes in each post are thoroughly reviewed. The raters use an open coding method~\cite{lune2017qualitative} to label SO posts and categorize them. Each SO post is reviewed by two raters. We use the \textit{“Google Sheets”} platform~\cite{googleSheet} to save all extracted labels in an online environment. That is, all raters put generated labels in the shared document, but they do not have access to the labels that other raters assigned to each post. After labeling SO posts in each round, the raters meet to discuss disagreements and resolve conflicts. In the case they fail to resolve a disagreement, a third rater reviews the SO post and makes a decision about its label, acting as a tie-breaker. Besides, the raters review the generated labels in each meeting to ensure their comprehensiveness and granularity (combining similar labels generated by different raters or dividing a label into separate ones). The justification for dividing SO posts into 10 labeling rounds, with approximately 10\% of the total collected posts in each round, emerges from the necessity for raters’ discussion in iterative meetings. These discussions are essential for resolving labeling conflicts and reaching an agreement, reviewing generated labels, and creating finer or coarser-grained labels.
In case of changing existing labels, raters re-review the previously labeled posts to ensure that assigned labels are in synchronization with newly generated labels. We made this decision to allow for continuous improvement of the labeling process. This way raters have the opportunity to resolve their conflicts at the end of each round, similar to the technique used in previous studies~\cite{humbatova2020taxonomy, islam2019comprehensive}. Besides, in the case that any of the raters suggests generating a new label, all raters meet to discuss and reach an agreement on that new label. After completing labeling all 927 posts, all raters meet to finalize the generated labels, categorize them, and create the taxonomy. Then, the first two authors review all of the labeled posts again to ensure that their assigned labels are in sync with the final generated taxonomy.
Regarding the posts in which the questioner asks more than one question belonging to multiple categories of challenges, we repeat that record in our dataset and assign a different label to each record. For instance, post \href{https://stackoverflow.com/questions/45382763}{\#45382763} has been identified to belong to two categories, comprehension and design problems. 
Although we could not report inter-rater agreement level due to the lack of prior defined categories, we calculate inter-rater agreement between the pair of raters who investigate each SO post after finalizing the labels using Cohen's Kappa~\cite{mcdonald2019reliability} and obtained an 86\% agreement level. Table~\ref{tab:lablel_process} presents detailed information on the labeling procedure.

\begin{table}[]
    \centering
    \caption{Detailed information on the manual labeling process.}
    \begin{tabular}{c r r r r}
        Round & Analyzed  SO posts & Conflicts & Relevant to DRL & New categories\\
        \hline
         1 & 100 & 11 & 95 & 15 \\
         2 & 100 & 7 &  93 & 1 \\
         3 & 72 & 10 & 67 & -- \\
         4 & 100 & 16 & 96 & 4 \\
         5 & 100 & 14 & 90 & 1 \\
         6 & 100 & 9 & 96 & -- \\
         7 & 100 & 5 & 95& -- \\
         8 & 100 & 7 & 93 & 1\\
         9 & 100 & 5 & 94 & -- \\
         10 & 55 & 3 & 51 & -- \\
         \hline
        \textbf{Total} & \textbf{927} & \textbf{87} & \textbf{870} & \textbf{22}\\
        \hline
    \end{tabular}
    \label{tab:lablel_process}
\end{table}

During the manual inspection of SO posts, we filtered out 57 posts that were not related to DRL development. Generally, some questioners may add DRL-related tags to their posts by mistake or because of unfamiliarity with DRL and its real
capabilities to solve their considered problems
(such as \href{https://stackoverflow.com/questions/60958362}{\#60958362}). We also filtered out posts that are too general and which could not be considered as reporting about a challenge in DRL development (e.g., \href{https://stackoverflow.com/questions/3972812}{\#3972812}). 

It is worth mentioning that 70\% of the DRL-related questions in SO still remain without any accepted answer.
This is consistent with previous findings by Alshangiti et al.~\cite{alshangiti2019developing} that 61\% of ML-related SO posts remain without any accepted answer.
Multiple factors could explain this finding. 
In some cases, the person who asks the question responds to the question after a while, but she does not assign the accepted answer badge to the post (e.g., \#\href{https://stackoverflow.com/questions/45364837}{45364837}). Some users also ask basic questions irrelevant to DRL but assign DRL-related tags to them. These questions receive negative scores and remain without any accepted answer (e.g., \#\href{https://stackoverflow.com/questions/50544568}{50544568}). We also observe posts where the person asking the question forgot to assign the accepted answer badge to an answer, based on upvotes to the response, comments of the person who asked the question, or other people with the same problem under the response (e.g., \#\href{https://stackoverflow.com/questions/63250935}{63250935}).  
About the posts with accepted answers, it should be mentioned that 16\% of them have been answered by the user who published the question. This usually happens when a user asks a very specific question that remains unanswered for a long time, and then the same user finds the response elsewhere and adds it to his original post (e.g., \#\href{https://stackoverflow.com/questions/2723999}{2723999}).

\subsection{Taxonomy Construction and Validation}
\label{sec:taxonomy_const}
Similar to previous studies~\cite{vijayaraghavan2003bug,humbatova2020taxonomy}, 
we use a bottom-up methodology to create the taxonomy.
In fact, after completing each labeling round, we categorize all generated labels belonging to a similar theme into a group. Next, we build up parent nodes in a way that makes sure that categories and their subcategories adhere to a \textit{`is a'} relationship. Concerning that the raters may provide new labels during each labeling round, we need to update the taxonomy which means adding a new category, a new subcategory, or combining two categories/subcategories. After any update on the taxonomy which leads to a new version of it, all the authors have a debate on the newly generated version of the taxonomy in a group meeting. By completing the final labeling round and integrating all updates on the taxonomy, all paper’s authors carry out a careful inspection of the produced taxonomy (including all categories and subcategories) over a meeting and finalize it. 

Interviews and surveys stand as the two popular methods to validate the results of qualitative studies~\cite{hove2005experiences,aldhaen2020interview}. Considering the advantages of conducting surveys, including cost-effectiveness, generalizability, reliability, and versatility~\cite{decarlo2018scientific,nekkanti2016surveys}, we assessed the comprehensiveness and representativeness of the obtained taxonomy using a survey with DRL developers/practitioners who are not involved in the construction of the taxonomy. Nevertheless, it is noteworthy that several preceding studies have presented their findings without undergoing any validation process~\cite{zhang2019empirical,islam2019comprehensive}.

While we build our taxonomy based on SO posts, we use Github to identify potential respondents to our survey. We collect a list of survey participants from collaborators of GitHub repositories related to DRL. Specifically, we extract GitHub repositories mentioning \textit{'deep reinforcement learning'} in their description using GitHub's search API V3; a rest API that receives a query and returns a list of repositories that satisfy conditions stated in the query. In other words, we use \textit{‘deep reinforcement learning’} as the keyword to generate the search query of GitHub search API V3\footnote{https://docs.github.com/en/rest}. Given that GitHub search API limits access to only the first 1,000 results, we follow the methodology used in~\cite{morovati2023bugs} and run several different queries to achieve less than 1,000 repositories for each query. That is, we divide the duration of the search for repository creation date between Jan 1, 2010, and Jan 31, 2023 (the date of running queries) into snapshots of 1 month. Thus, we execute 157 GitHub search requests to collect 7,244 repositories. Subsequently, we filter out forked and disabled repositories. The complete list of repositories and a sample search query to extract DRL-related repositories are accessible via the replication package~\cite{replication_package}. In the next step, we check the repositories’ contributors and collect the contributors mentioning their email addresses, obtaining 2531 unique email addresses of developers.

We use Qualtrics~\cite{survey_ualtrics}, an online survey tool for designing and conducting surveys, to create survey forms.
Table~\ref{tbl:surveyStructure} presents the structure of our survey questionnaires. The survey starts with general questions regarding the participant's current role and experience in DRL development (Section 1 of Table~\ref{tbl:surveyStructure}). Subsequently, we delve into specific questions regarding each DRL development challenge which is pointed out in the finalized taxonomy (Section 2 of Table~\ref{tbl:surveyStructure}). Concerning the potential complexity and difficulty of comprehending the whole taxonomy as a single figure within the survey, we present challenges (subcategories) in groups based on their respective main categories. Besides, to ensure clarity, a detailed description is given with each challenge offering participants a thorough understanding of the asked challenges. 
For each identified challenge, we ask three questions including 1) a \textit{`yes/no'} question identifying whether the answerer has faced the identified challenge, 2) the severity of the challenge, and 3) the amount of effort required to address the challenge. If a participant answers \textit{`yes'}  to the first question, s/he will have the next two Likert-scale questions which are related to the severity of challenges and the required effort to address them. We also provide a free-text question in the final part of the survey asking the participants about any challenges in developing DRL applications not listed in our provided taxonomy (Section 3 of Table ~\ref{tbl:surveyStructure}). These free-text questions allow us to collect possible challenges that we may miss in the taxonomy. The full survey questionnaire is available in our replication package~\cite{replication_package}.

We conduct a comprehensive analysis of top DRL-related GitHub repositories to ensure the completeness of the generated taxonomy. Besides, investigating the challenges faced by developers engaged in the development of real-world DRL applications may further enhance the generalizability of the generated taxonomy. Our methodology for selecting DRL-related repositories aligns with established approaches documented in similar prior studies~\cite{morovati2023bugs,morovati2024bug,humbatova2020taxonomy}. Initially, we extracted 7,244 repositories to identify contributors of DRL-related repositories. Then, we select the top 100 repositories based on the highest number of stars. Subsequently, we conduct searches for `challenge', `difficult', and `complex' keywords within repositories’ documentation, commit messages, and closed issues to identify potential challenges encountered by developers during the development of these repositories. Our search yielded 746 occurrences of the aforementioned keywords within the top 100 DRL-related repositories. Using a confidence level of 95\% and a confidence interval of 5\%, we conduct sampling, resulting in the selection of 254 instances. In the next step, two raters independently examine 254 randomly selected commits and issues to find out whether they pertained to real DRL-related development challenges. The raters carefully review commit messages, issue discussions, and investigate any changes made in relation to the specified keywords. 

\begin{table}[]
    \centering
    \caption{Survey structure}
    \label{tbl:surveyStructure}
    \resizebox{0.55\columnwidth}{!}{
        \begin{sideways}
        \begin{tabular}{p{1cm} l |l l | l l l| l l l}
            \hline
            \textbf{Sec.1} &\multicolumn{9}{l}{\textbf{General questions}}\\
             & \multicolumn{3}{l}{\textit{Email address (optional)}}&
            \multicolumn{6}{l}{\textit{\framebox(50,10){}}}\\
             & \multicolumn{3}{l}{\textit{Years of experience on DRL }} &
            \multicolumn{6}{l}{\textit{\framebox(50,10){}}}\\
             & \multicolumn{9}{l}{\textit{\textbf{\dots}}}\\
            \hline
            \textbf{Sec.2}& \multicolumn{9}{l}{\textbf{DRL development challenges review}}\\
            & \multicolumn{1}{l}{\textbf{\textit{DRL Challenges}}} & \multicolumn{2}{l}{\textbf{\textit{Facing this challenge}}} & \multicolumn{3}{c}{\textbf{\textit{Challenge severity}}} & \multicolumn{3}{c}{\textbf{\textit{Needed effort to address}}} \\
            & \textit{Comprehension}& \multicolumn{1}{c}{$\bigcirc$\textit{Yes} } & \multicolumn{1}{c|}{$\bigcirc$\textit{No}}
             & $\bigcirc$\textit{Minor} &$\bigcirc$\textit{Major} &$\bigcirc$\textit{Critical} &$\bigcirc$\textit{Low} &$\bigcirc$\textit{Medium} &$\bigcirc$\textit{High} \\
             & \textit{Reward}& \multicolumn{1}{c}{$\bigcirc$\textit{Yes} } & \multicolumn{1}{c|}{$\bigcirc$\textit{No}}
             & $\bigcirc$\textit{Minor} &$\bigcirc$\textit{Major} &$\bigcirc$\textit{Critical} &$\bigcirc$\textit{Low} &$\bigcirc$\textit{Medium} &$\bigcirc$\textit{High} \\
             & \multicolumn{8}{l}{\textit{\textbf{\dots}}}\\
            \hline
            \textbf{Sec.3} & \multicolumn{9}{l}{\textbf{Free-text question}}\\
             & \multicolumn{9}{l}{\textit{Any challenge that we have missed}}\\
             & \multicolumn{9}{l}{\framebox(200,20){}}\\
             & \multicolumn{9}{l}{\textit{Any suggestion or comment}}\\
             & \multicolumn{9}{l}{\framebox(200,20){}}\\
            \hline
        \end{tabular}
        \end{sideways}
    }
\end{table}

\section{Results 
}
\label{sec:results} 
This section presents and discusses the results of our study. All the materials used in this study, including the collected data, are publicly available online in our replication package at~\cite{replication_package}. 

\subsection*{\textit{RQ1:} What are the common challenges of DRL application development?}

\begin{figure*}
    \centering
\includegraphics[scale=0.65,angle=90]{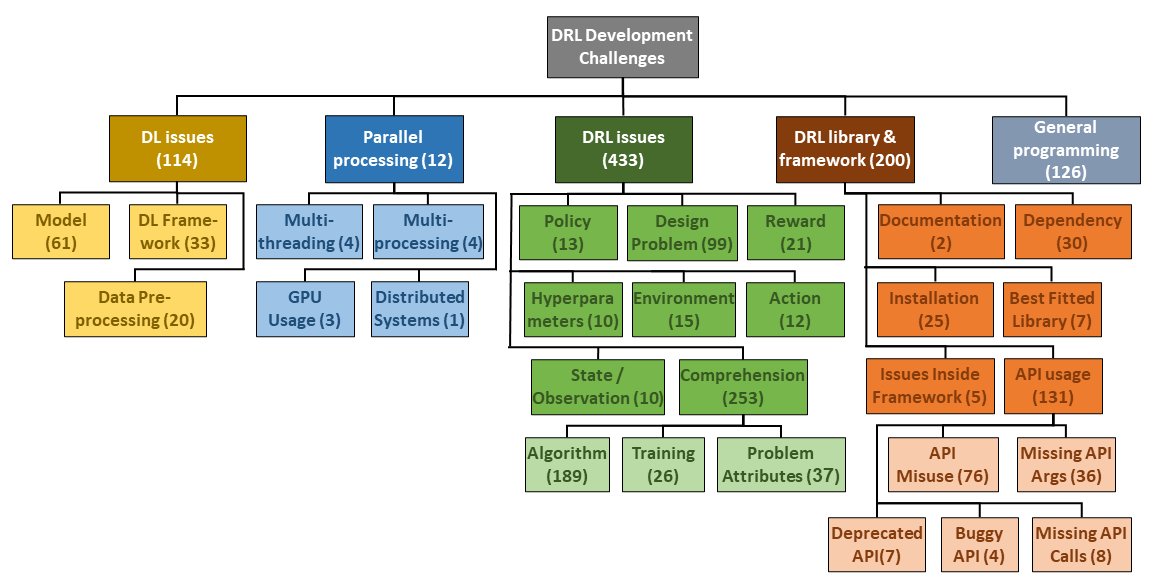}
    \caption{Taxonomy of common challenges in DRL development. Numbers represented with each category/subcategory indicate the number of SO posts categorized into that category/subcategory.}
    \label{fig:taxonomy}
    \vspace{-1em}
\end{figure*}

The final taxonomy of challenges of DRL application development is arranged into a tree structure with five high-level categories in which leaves (subcategories) 
refer to the challenges. Fig.\ref{fig:taxonomy} shows the taxonomy of common challenges in the development of DRL applications. 
The number in parentheses presented next to each category/subcategory is the absolute number 
of identified SO posts that are categorized into that category (the absolute frequency 
of each challenge). 
To give a better understanding of identified challenges, 
a brief description of each category/subcategory is provided in the following. 

\textbf{DRL Issues.} 
This category focuses on the challenges that developers may face while developing the DRL part of their applications. Challenges that belong to this category are specific to DRL application development. That is, compared to challenges classified under other categories that can be shared among other kinds of ML-related applications (e.g. supervised learning), challenges of the \textit{DRL issues} category are only faced during DRL application development. 

\begin{enumerate}[label=\alph*.]
    \item \emph{Design Problem:} Instances where the user asks advice for designing a solution and implementing a DRL application for their specific problem or scenario. For example, developers asked for recommendations to implement different parts of DRL applications for Curve Fever or mini-golf games. Another such challenge in this category is related to designing the properties of each object in a tank game.  
    Although we tried to make finer-grained subcategories under this subcategory, we found out that it is impractical to split this subcategory further, as the majority of this subcategory's posts primarily comprised users seeking guidance on high-level conceptual queries about defining their problem within a DRL context.
    \item \emph{Comprehension:} Challenges about the meaning or details of theoretical concepts in DRL, i.e., misunderstanding about basic formulas of different DRL algorithms. For instance, a developer mentioned \textit{\enquote{I'm trying to make a learning football game from scratch using Deep Q-learning algorithm (without convolutional network though). I just couldn't figure out what does $\Phi$ stand for in this algorithm.}}, or one which is related to the difference between SARSA and Q-learning algorithms in terms of collecting the next policy value.
    \begin{enumerate}[label*=\arabic*.]
        \item \emph{Training:} 
        This subcategory comprises inquiries concerning the theoretical concepts of the DL model and its significance in DRL applications. The DL model plays an essential role in the training and decision-making process of the DRL agents. Indeed, DL models in DRL applications serve to represent policy of the agent, estimate the value of state-action, and learn mappings from states to values.
        \item \emph{Problem attributes:} 
        SO posts containing theoretical queries about attributes of DRL problems (such as reward, action, state, etc.) are categorized in this subcategory. SO posts in this category are related to the conceptual aspects of reward, action, and state, not their implementation details. For example, questions on how to formulate states, actions, and the reward signal for a particular DRL problem, or why a particular definition of the states is not suitable for a problem.
        \item \emph{Algorithm:} 
        Questions related to fundamental concepts of various DRL algorithms, such as actor-critic, Q-learning, SARSA, etc., are categorized as algorithm subcategories. The main duty of DRL algorithms is updating parameters of the policy and value function according to the observed states and achieved rewards. Our investigation into these algorithms reveals that Q-learning (28\%) and DQN (15.3\%) are the most common DRL-related algorithms posing challenges for developers. Conversely, Actor-Critic (3.1\%), Proximal Policy Optimization (2.6\%), and Trust Region Policy Optimization (0.5\%) are the least queried DRL-related algorithms.
    \end{enumerate}
    \item \emph{Policy's Loss:} This category refers to challenges about DRL learning policy's loss. As an example, questions regarding implementing a customized loss function or any problems in loss calculation methods are categorized in this group. 
    \item \emph{Reward:} Challenges in the implementation of reward, e.g., not using the negative reward to penalize each added time step are categorized in this subcategory. An example of this category is a post asking \textit{\enquote{I am implementing the basic RL algorithm to play the game Flappy Bird. I want to be able to process the screen and recognize whether a point has been scored or the bird has died. Processing the screen returns a stacked numpy array. The reward function then needs to assign a reward to the provided array, but I have no idea how to go about this}}.
    \item \emph{Action:} Questions/Challenges related to the action(s). e.g., possible actions in a specific game or implementing a \textit{`chooseAction'} method for a PacMan bot. 
    \item \emph{State/Observation:} Questions/Problems regarding the state(s) or the agent observation(s), e.g., handling large state spaces. An example of such instance is a user asking \emph{\enquote{I implemented a 3x3 OX game by q-learning (it works perfectly in AI v.s AI and AI v.s Human), but I can't go one step further to 4x4 OX game since it will eat up all my PC memory and crash\ldots Since I need to calculate each Q value (for each state, each action), I need such a large number of array, is it expected? any way to avoid it?}} and the accepted answer offered suggestions on reducing their state space size by considering symmetries and other tricks.
    \item \emph{Environment:} Questions/Problems pertinent to the environment, e.g., designing a custom environment. For example a user asked \emph{\enquote{I'm very new to Ray RLlib and have an issue with using a custom simulator my team made. We're trying to integrate a custom Python-based simulator into Ray RLlib to do a single-agent DQN training. However, I'm uncertain about how to integrate the simulator into RLlib as an environment}}. 
    \item \emph{Hyperparameters:} Questions/Challenges related to the hyperparameters of the RL algorithm, e.g., setting the discount factor too high. A good example of this group is demonstrated in a post where the user shared the code for the learning algorithm and reported that the loss keeps increasing and the model is not learning. The accepted answer states that \emph{\enquote{The main problem I think is the discount factor, gamma. You are setting it to 1.0, which means that you are giving the same weight to the future rewards as the current one}}. 
\end{enumerate}

\textbf{DRL Libraries/Frameworks.} This category refers to the challenges that developers face when they are trying to use DRL-specific libraries/frameworks (e.g., KerasRL~\cite{plappert2016kerasrl}, RLlib~\cite{liang2018rllib}, Tensorforce~\cite{lift-tensorforce}, etc.). 
Challenges that software developers face when using libraries/frameworks have been extensively studied for traditional software systems development~\cite{decan2019empirical,nguyen2010graph} and also for DL applications development~\cite{arpteg2018software}. However, despite the large number of SO posts related to the usage of DRL libraries/frameworks (i.e., 200 SO posts), these challenges are yet to be examined for DRL application development. 

\begin{enumerate}[label=\alph*.]
    \item \emph{Installation:} Questions/problems regarding installing/uninstalling DRL-related libraries/frameworks or missing libraries.
    Issues categorized in this subcategory can often stem from an incompatibility between DRL-related frameworks/libraries and other libraries. 
    For example, a user described her issue as \emph{\enquote{when I try to install gym[box2d] I get the following error: I tried: pip install gym[box2d]. on anaconda prompt I installed swig and gym[box2d] but I code in python3.9 env and it still not working (my text editor is pycharm) gym is already installed}.}.
    \item \emph{Dependency:}
    This subcategory includes questions/challenges about the mismatch between versions of installed libraries/frameworks and problems in installed versions of libraries, e.g., when the version of the installed \emph{OpenAI Gym} is not compatible with \emph{Python}. An instance of this subcategory is a user reporting getting an error while installing OpenAIGym and the answer pointed out that \emph{\enquote{the error means that the package has dependency requirements that conflict with one another}}. 
    \item \emph{API usage:} This subcategory includes questions about the usage of arguments, attributes, methods, etc. of an API. It also includes questions about the default values, implemented method, existence of attributes, or methods in an API. An example from this group of issues is a user reporting not knowing how to get the weights of the network using the correct API methods: \emph{\enquote{I'm using RLlib to train a reinforcement learning policy (PPO algorithm). I want to see the weights in the neural network underlying the policy. After digging through RLlib's PPO object, I found the TensorFlow Graph object. I thought that I would find the weights of the neural network there. But I can't find them}}. 
    This subcategory is subdivided into five subcategories to delineate more detailed challenges. 
    \begin{enumerate}[label*=\arabic*.]
        \item \emph{API misuse:} This subcategory covers SO posts that mention misunderstanding of API usage. In other words, API misuse occurs when DRL developers try to utilize an API in a manner that is not aligned with its intended purpose. 
        \item \emph{Missing API call:} Questions related to the absence of necessary API calls within a code snippet are classified in this subcategory.
        \item \emph{Missing API args:} When SO posts discuss challenges that DRL developers face due to the absence of one or more essential arguments in an API call, we classify them under this subcategory.
        \item \emph{Buggy API:} This subcategory includes SO posts that inquire about calling APIs resulting from the bugs within the implementation of APIs. It is worth mentioning that this challenge is distinct from issues related to the implementation of DRL applications themselves; rather, they pertain to the implementation of the API. 
        \item \emph{Deprecated API:} In this subcategory, we cover questions about calling a deprecated API which has been altered or removed from the library/framework.
    \end{enumerate}
    \item \emph{Documentation (using newly added features):} This subcategory of issues occurs when a developer wants to use a feature of a DRL library/framework, but there is no documentation for it. For example, a user who could not find the required documentation for the \emph{Neural Network Approximator} in \emph{ReinforcementLearning.jl}: \emph{\enquote{I have decided to use a Neural Network Approximator. But the docs do not discuss much about it, nor are there any examples where a neural network approximator is used. I am stuck on how to figure out how to use such an approximator}}. 
    \item \emph{Best fitted library for a special task (library suitability):} Instances where the user asks about the best DRL libraries/frameworks for customizing agents, based on the requirements of the problem. An instance of this group was observed in a post where a user had a customized state space and was looking for a library that supports it: \emph{\enquote{I've had some luck training an agent using keras-rl, specifically the DQNAgent, however, keras-rl is under-supported and very poorly documented. Any recommendations for RL packages that can handle this type of observation space? It doesn't appear that openai baselines, nor stable-baselines can handle it at present}}. 
    \item \emph{Problems inside DRL frameworks:} Including issues that are encountered because of internal faults, i.e., bugs in the DRL frameworks. For example, a user kept getting a \emph{numpy} error when calling the \emph{model.learn()} function and it was found to be an official bug in the used library \emph{Stable Baselines3}.
\end{enumerate}

\textbf{DL Issues.}
This category represents the challenges that arise specifically from the DL part of DRL applications. As the challenges belonging to this category are shared by both DL and DRL applications, we use the high-level categories of the taxonomy provided by Humbatova et al.~\cite{humbatova2020taxonomy} for DL applications.
\begin{enumerate} [label = \alph*.]
    \item \emph{Model:} Questions regarding the DL model including model layer, activation function, load/save model, etc. For instance, a user who was implementing a DRL model asked for advice on back propagation: \emph{\enquote{I am struggling with the implementation of the back propagation. Since the rewards are so big, the error values are huge, which creates huge weights. After a few training rounds, the weights to the hidden layer are so big, my nodes in the hidden layer are only creating the values -1 or 1}}.
    \item \emph{Data Preprocessing:} Questions about preparing data to be fed into the DL model, e.g., the shape of the input matrix. As an example, there was a user who asked \emph{\enquote{I am learning how to use Gym environments to train deep learning models built with TFLearn. At the moment my array of observations has the following shape: (210, 160, 3). Any recommendations on what is the best way to reshape this array so it can be used in a TensorFlow classification model?}} 
    \item \emph{DL framework:} Questions about the usage of DL frameworks (e.g., Keras, TensorFlow, PyTorch, etc.) in development of DL part in DRL applications. For instance, a user who wanted to use \emph{Huber loss} in their model which was written using \emph{Keras}, but this loss function is not readily available in \emph{Keras}, and the accepted answer implements the \emph{Huber loss} function.
\end{enumerate}

\textbf{Parallel Processing \& Multi-threading.}
This category focuses on the challenges associated with running DRL applications as parallel or distributed applications. While running different types of DL applications in a parallel style is a common practice (e.g., using GPU or multi-core CPU), it is important to note that the architecture of DRL applications differs from other types of DL applications. As a result, running a DRL application as a parallel or distributed application introduces new challenges that may not occur in DL application development.
\begin{enumerate} [label = \alph*.]
    \item \emph{GPU usage:} Questions/Problems regarding utilizing GPU for running DRL applications. For example, a user was facing performance issues when training a DQN on GPU: \emph{\enquote{I am try to train a DQN model with the following code. The GPU (cuda) usage is always lower than 25 percent. I know the tensorflow backend is consulting the GPU resources, but the usage is low. Is there any way I can improve the utilization of the GPU (When I train a CNN network, the GPU (cude) utilization is around 70 percent)?}}
    \item \emph{Distributed processing:} Questions/Challenges about running DRL applications as a distributed software. For example when a user wanted to implement the \emph{Asynchronous Advantage Actor Critic (A3C)} model for reinforcement learning in the local machine, she posed a question about the possibility of implementing it in a distributed manner: \emph{\enquote{Would it be easier/faster/better to implement this using the distributed TensorFlow API? In the documentation and talks, they always make explicit mention of using it in multi-device environments. I don't know if it's an overkill to use it in a local async algorithm}}.
    \item \emph{Multi-threading:} Questions/Problems about running DRL applications as a multi-threaded software. An example of this category was observed in a post where the user asked about the possibility of reducing the DRL application's training time by running it on multiple threads concurrently: \emph{\enquote{My friend and I are training a DDQN for learning 2D soccer. I trained the model about 40.000 episodes but it tooks 6 days. Is there a way for training this model concurrently? For example, I have 4 core and 4 thread and each thread trains the model 10.000 times concurrently. Therefore, time to training 40.000 episodes are reduced 6 days to 1,5 days like parallelism of for loop}}.
    \item \emph{Multi-processing:}
    This subcategory refers to challenges stemming from running DRL applications 
    on multiple processing units (e.g., multiple CPUs). As an example, a user asked a question about using Ray\footnote{https://www.ray.io/} in a multi-processing style.
\end{enumerate}

\textbf{General Programming Issues.}
This category contains programming and coding mistakes occurring when developing DRL applications. 
The challenges in this category could not be classified in any of the other categories described above. For example, a user had a question about how to slice a 3D \emph{numpy} array in an RL finite MDP application (\href{https://stackoverflow.com/questions/67089715}{\#67089715}). \\ 

\begin{table}[]
    \centering
    \caption{Detailed information regarding tags and keywords used to extract DRL-related SO posts. Columns show the number (\#) and percentage (\%) of SO posts categorized in the main categories having mentioned tags/keywords (`DRL', `DRL LF', `DL', `PP', and `Gen' refer to `DRL issues', `DRL libraries/frameworks', `DL issues', `Parallel processing', and `General programming', respectively). 
    }
    \resizebox{0.95\columnwidth}{!}{
        \begin{tabular}{c l r r r r r r}
          \toprule
                 & \textbf{\makecell[b]{Tag/\\Keyword}} & 
                 \textbf{\makecell[b]{Posts\\(\#)}} & \textbf{\makecell[b]{DRL \\ (\%)}} & \textbf{\makecell[b]{DRL \\LF(\%)}} &\textbf{\makecell[b]{DL \\(\%)}} & \textbf{\makecell[b]{PP\\(\%)}} & \textbf{\makecell[b]{Gen\\(\%)}} \\
            \midrule
            \multirow{14}{*}{\rotatebox[origin=c]{90}{\textbf{Tags}}} & reinforcement-learning & 710 & 96& 58 & 86.8 & 83.3 & 54.8 \\
            & openai-gym & 208 & 10 & 59 & 10.5 & 25 & 25.4 \\
            & tensorflow & 159 & 11.3 & 14 & 47.4 & 16.7 & 20.6 \\
            & q-learning & 137 & 25.2 & 2 & 13.2 & - & 7.1 \\
            & deep-learning & 127 & 13.4 & 6.5 & 32.5 & - & 15.1 \\
            & keras & 78 & 5.1 & 6 & 29.8 & - & 7.9 \\
            & pytorch & 65 & 6.5 & 3 & 18.4 & - & 7.9 \\
            & dqn & 39 & 4.4 & 2 & 7 & 8.3 & 5.6 \\
            & stable-baselines & 24 & - & 5.5 & 1.8 & - & 5.6 \\
            & keras-rl & 19 & - & 3.5 & 5.3 & - & 2.4 \\
            & tensorflow-agents & 12 & 0.5 & 2.5 & 2.6 & - & 1.6 \\
            & rllib & 9 & 0.5 & 4 & - & - & - \\
            & tensorboard & 8 & - & 2 & - & - & 2.4 \\
            & starcraftgym & 5 & - & 2 & - & - & 0.8 \\
            \hline
            \multirow{14}{*}{\rotatebox[origin=c]{90}{\textbf{keywords}}} & reinforcement learning & 313 & 44.8 & 24 & 36 & 41.7 & 19.8 \\
            & deep & 136 & 16.4 & 10.5 & 23.7 & 16.7 & 11.9 \\
            & ppo & 117 & 11.5 & 18 & 11.4 & 8.3 & 13.5 \\
            & \makecell[l]{deep reinforcement \\learning} & 37 & 4.6 & 2.5 & 5.3 & 8.3 & 4 \\
            & a3c & 17 & 1.6 & 0.5 & 3.5 & 25 & 1.6 \\
            & deep q-learning & 16 & 2.5 & 1 & 0.9 & - & 1.6 \\
            & actor critic & 11 & 1 & - & 4.4 & 8.3 & 0.8 \\
            & ddqn & 9 & - & 1 & 1.8 & 8.3 & -- \\
            & drl & 5 & 0.7 & 0.5 & - & - & 0.8 \\
            & dql & 5 & 0.7 & 0.5 & 0.9 & - & - \\
            & deep rl & 2 & 0.5 & - & - & - & - \\
            & trpo & 2 & 0.5 & - & - & - & - \\
            & hierarchical & 1 & 0.2 & - & - & - & - \\
            & \makecell[l]{trust region\\policy optimization} & 1 & 0.2 & - & - & - & - \\
            \bottomrule
        \end{tabular}
    }
    \label{tab:tag_keywords}
\end{table}

Overall, the majority of the analyzed SO posts have been assigned to categories \textit{Comprehension}, 
\textit{Design problem}, \textit{Model}, and \textit{API Usage}. 
Aside from \textit{Design problem} related questions, which are often quite specific (i.e., related to particular implementations of DRL), the majority of questions asked by DRL developers concern issues that apply to DRL applications in general. 

\begin{tcolorbox}[colback=blue!5,colframe=blue!40!black]
\textbf{Finding 1:} 
The taxonomy of challenges in DRL development is structured into five main categories including \textit{DRL issues}, \textit{DL issues}, \textit{DRL libraries/frameworks}, \textit{parallel processing \& Multi-threading}, and \textit{general programming issues}, where \textit{DRL issues} ($48.9\%$) and \textit{DRL libraries/frameworks} ($22.6\%$) categories are the most popular. Among the challenges, \textit{comprehension} ($28.6\%$), \textit{API usage} ($14.8\%$), and \textit{design problem} ($11.2\%$) are the most prevalent DRL development challenges. 
\end{tcolorbox}

\begin{figure}
    \centering
    \begin{subfigure}[t]{0.48\textwidth}
    \includegraphics[width=0.9\linewidth]{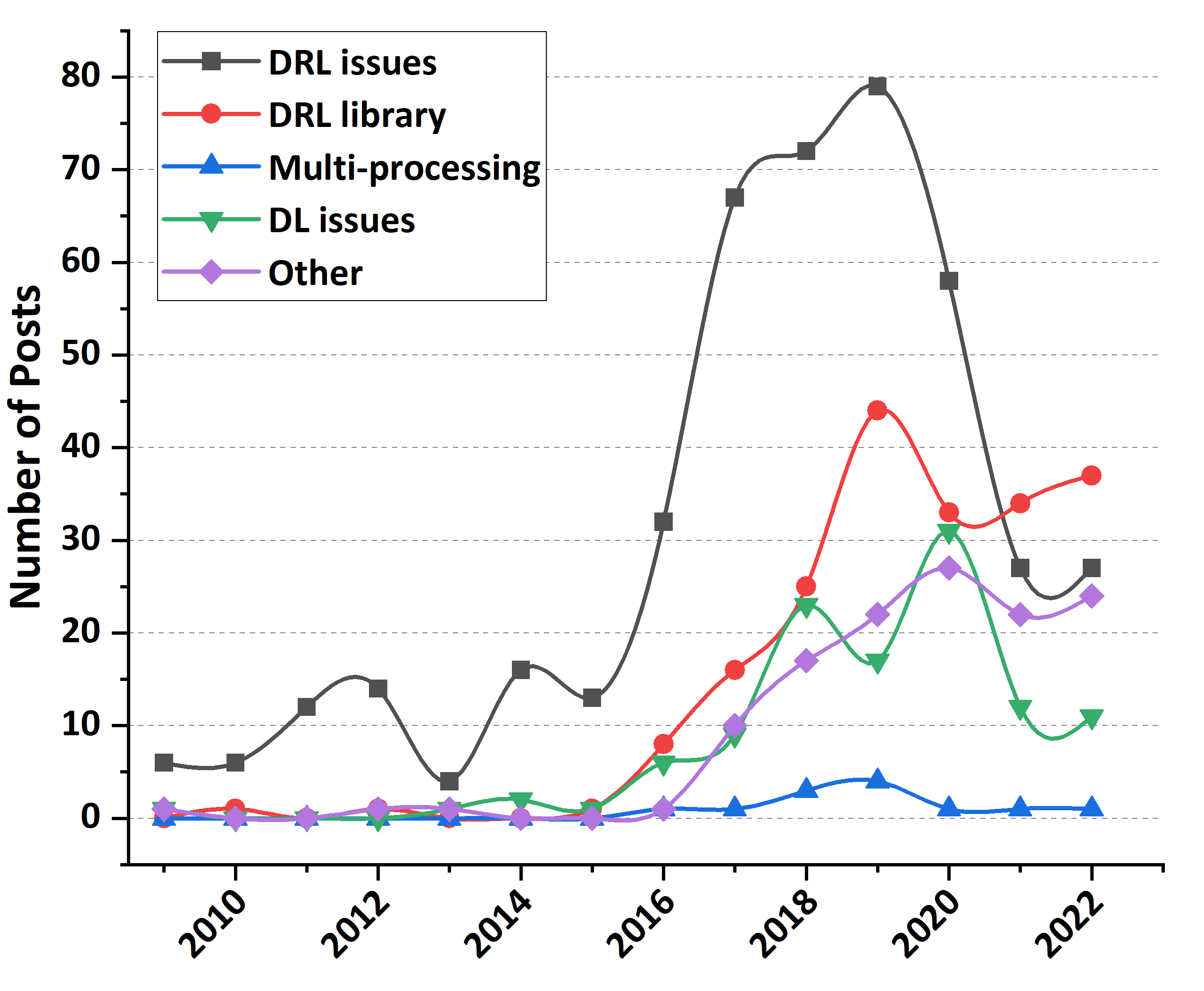}
    \caption{}
    \label{sub_fig:all_posts_per_year}
\end{subfigure}
\begin{subfigure}[t]{0.48\textwidth}
    \includegraphics[width=0.9\textwidth]{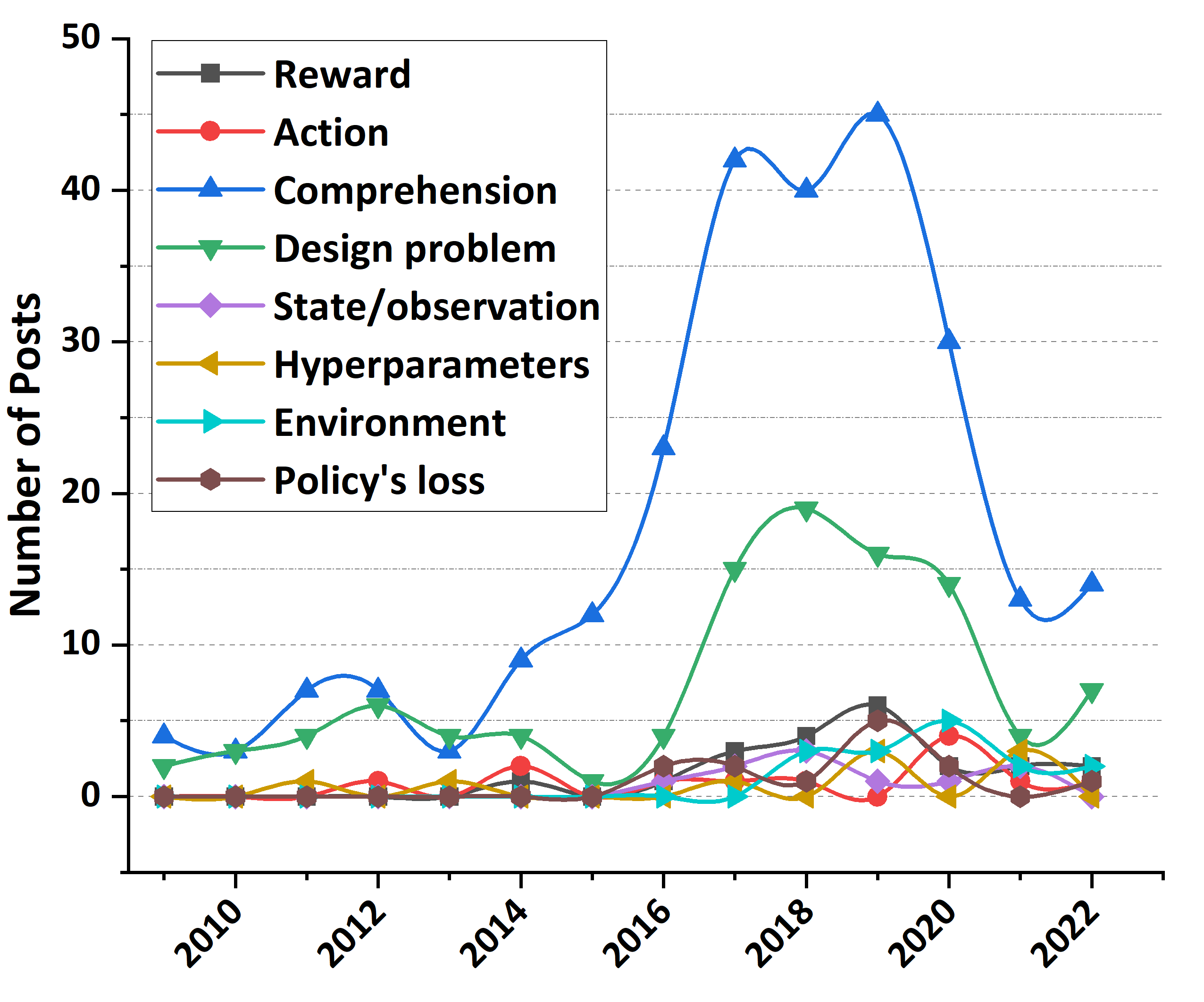}
    \caption{}
    \label{sub_fig:drl_posts_per_year}
\end{subfigure}\hspace{\fill} 
    \caption{Distribution of DRL related SO posts per years (a) high-level categories of the provided taxonomy, (b) subcategories of \textit{`DRL issues'} category.}
    \label{fig:posts_per_year}
\end{figure}

Table~\ref{tab:tag_keywords} shows the number and percentage of the SO posts within our dataset having different tags and/or keywords that we used in subsection \ref{subsec:so_extract}. 
Fig.\ref{fig:posts_per_year} presents the distribution of SO posts related to DRL application development for a period of 13 years, from 2009 to 2022. From  Fig.\ref{sub_fig:all_posts_per_year}, we observe a substantial surge of inquiries about DRL development in 2016; reaching a peak in 2019 and 2020. Additionally, as can be seen on Fig.\ref{sub_fig:drl_posts_per_year}, \textit{comprehension} and \textit{design problem} questions dominated posts about DRL application development challenges.  
It is also noticeable that \textit{API usage}, the second most common DRL application development challenge, was at its peak in 2018.

\begin{figure}[!h]
    \centering
    \includegraphics[width=0.70\columnwidth]{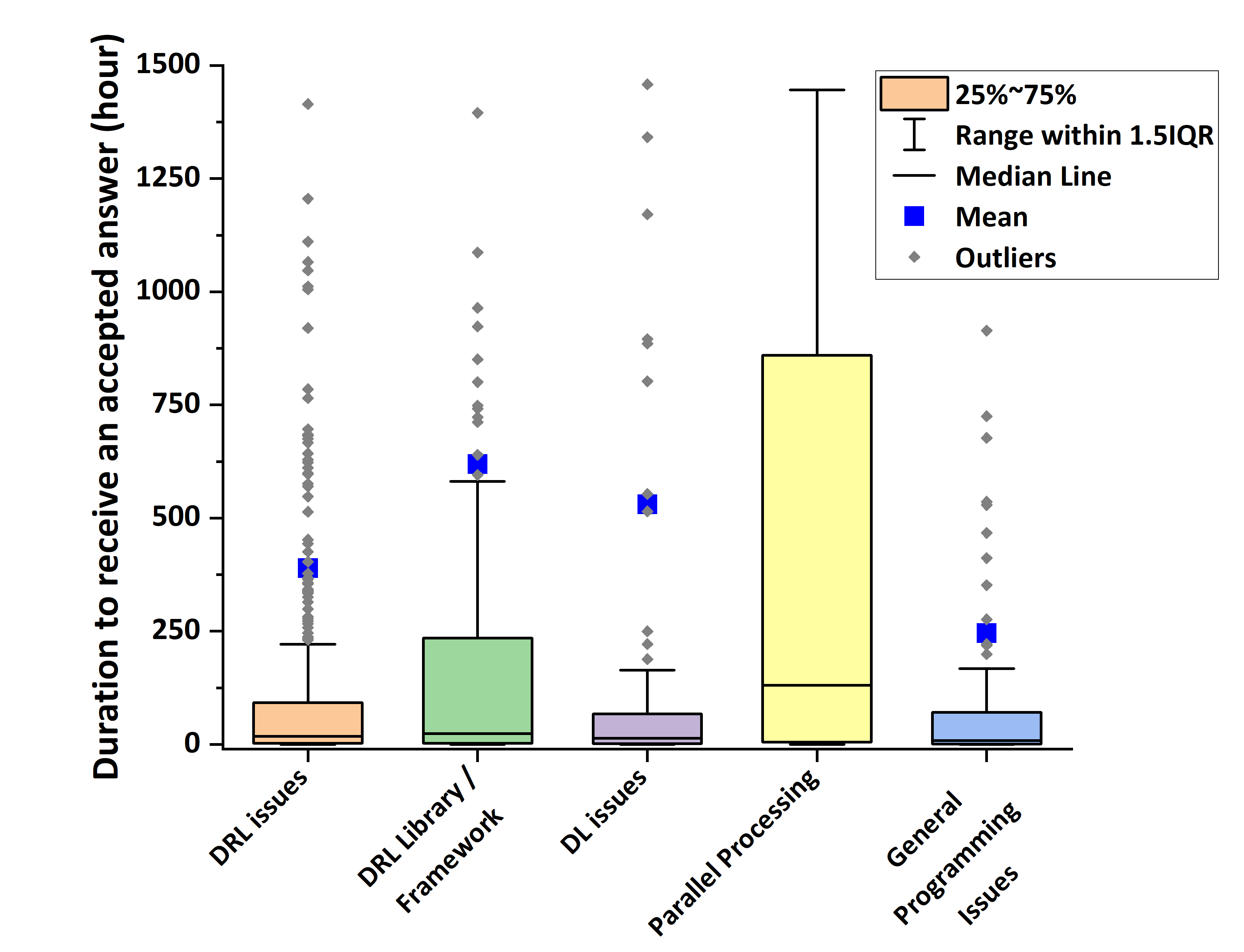}
    \caption{Duration to receive an accepted answer (hours).
    }
    \label{fig:post_duration}
\end{figure}

Fig.\ref{fig:post_duration} depicts the distribution of time taken by SO posts from different categories 
to receive an accepted answer.
This duration is an indicator of the difficulty level of the questions mentioned in the SO posts
in the development of both traditional ~\cite{haque2020challenges,zahedi2020mining} and ML software~\cite{alshangiti2019developing,chen2020comprehensive}.
\textit{Parallel Processing} is the category with the highest average time taken before receiving an accepted answer. This can be explained by the fact that using multi-processing or distributed processing in DRL is not necessarily widespread and also requires particular knowledge and expertise. The remaining categories need a nearly similar average time frame to receive an accepted answer, with the \textit{general programming issues} category having the shortest average duration. We attribute the shortest average time of the 
\textit{general programming issues} category to the fact that this category contains generic challenges that do not require expertise in DL or DRL. 

\begin{figure*}[!h]
\centering
\begin{subfigure}[t]{0.45\textwidth}
    \includegraphics[width=0.9\textwidth]{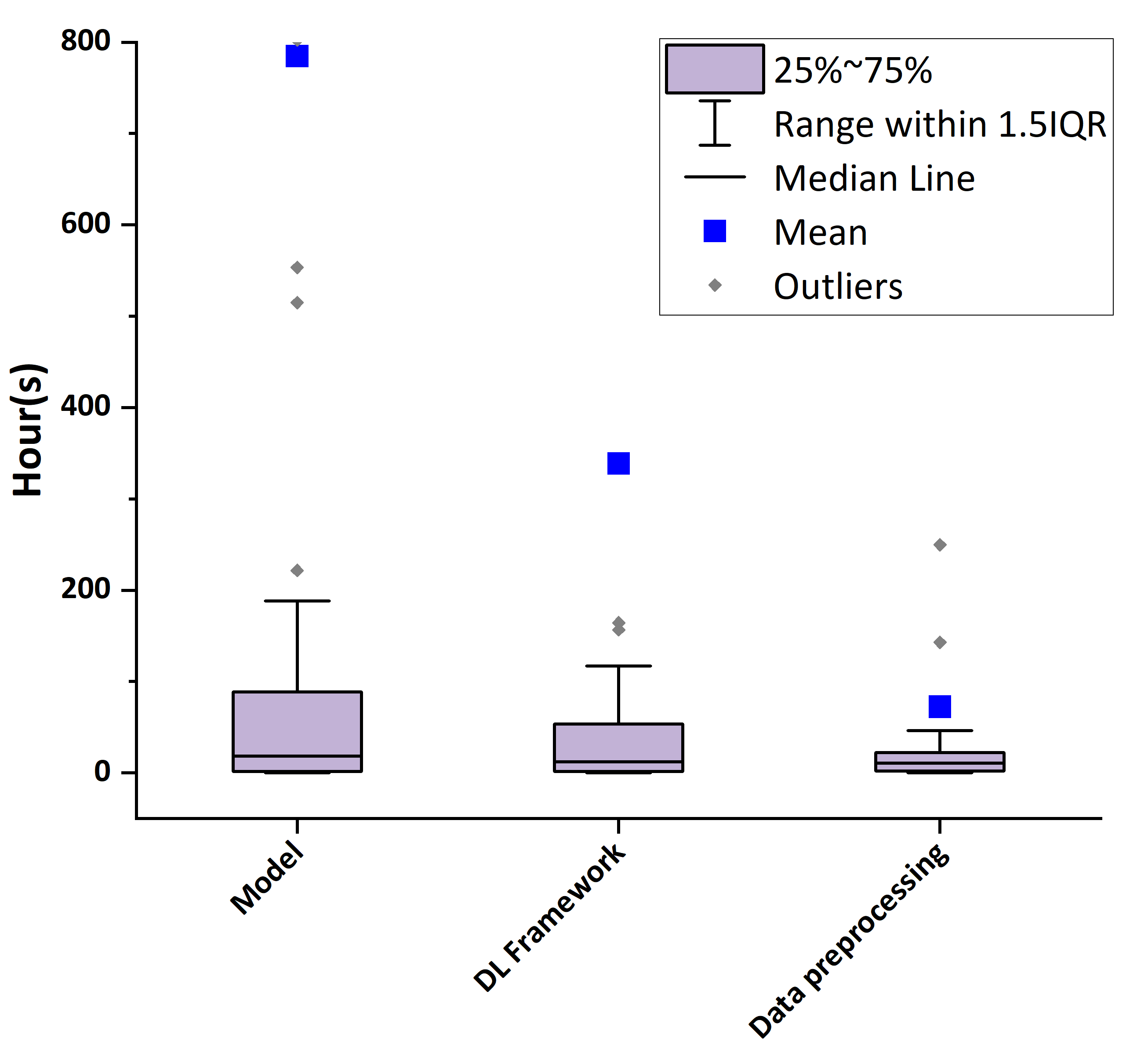}
    \caption{DL issues}
    \label{subfig:dl_issues_duration}
\end{subfigure}\hspace{\fill} 
\begin{subfigure}[t]{0.45\textwidth}
    \includegraphics[width=0.9\linewidth]{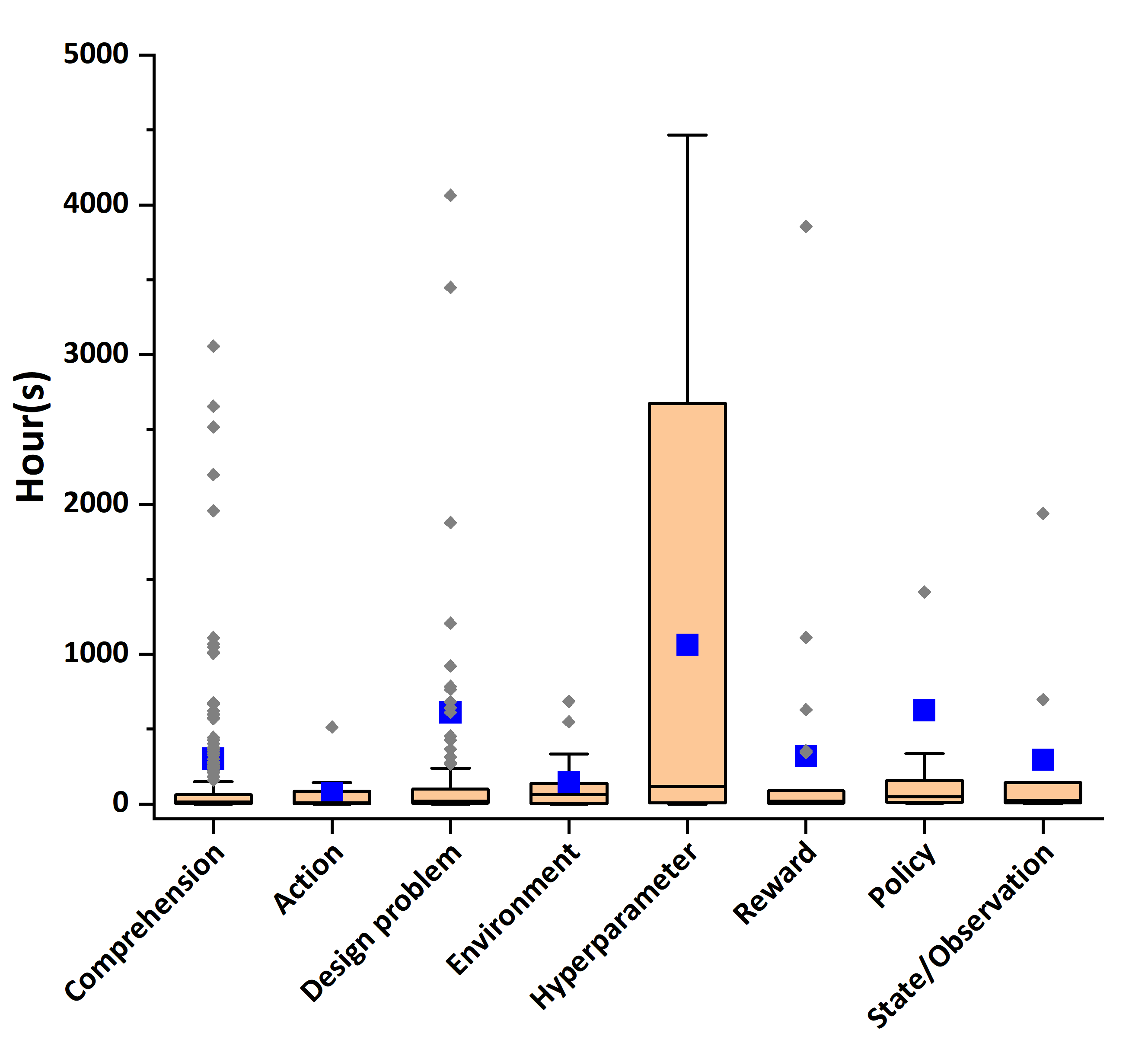}
    \caption{DRL issues}
    \label{subfig:drl_issues_duration}
\end{subfigure}

\bigskip 
\begin{subfigure}[t]{0.45\textwidth}
    \includegraphics[width=0.9\linewidth]{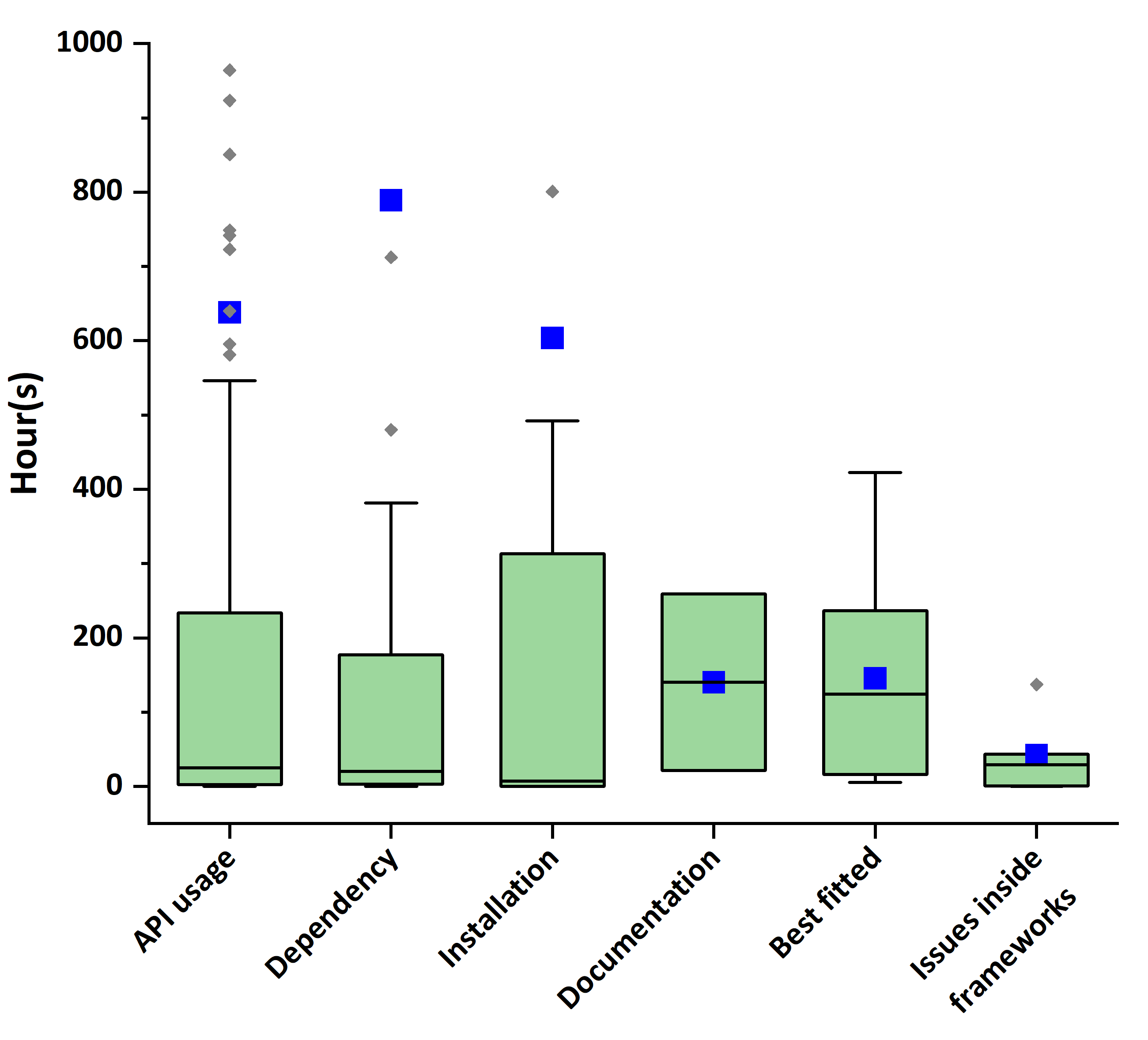}
    \caption{DRL libraries/frameworks}
    \label{subfig:drl_lib_duration}
\end{subfigure}\hspace{\fill} 
\begin{subfigure}[t]{0.45\textwidth}
\includegraphics[width=0.9\linewidth]{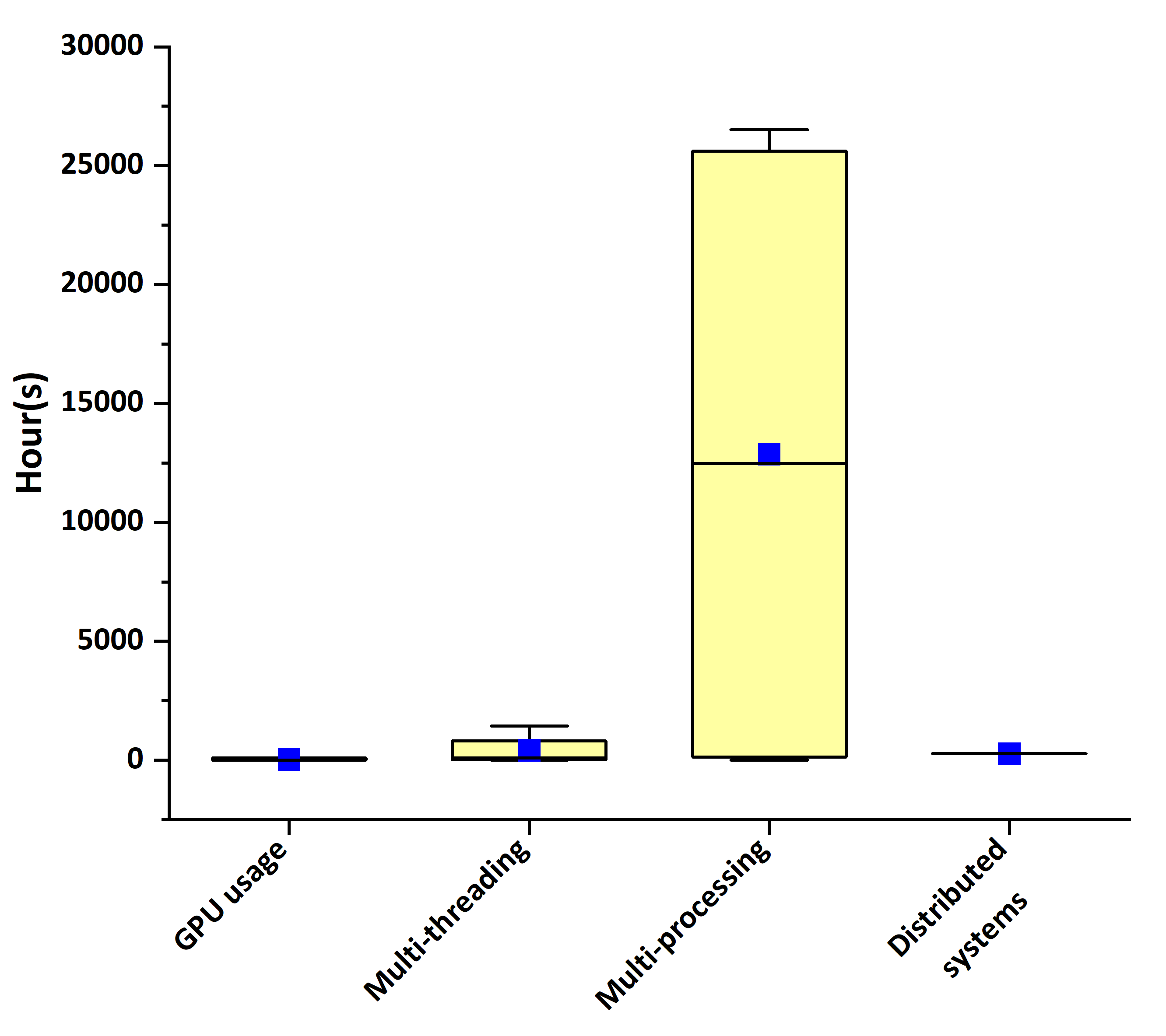}
    \caption{Parallel processing \& multi-threading}
    \label{subfig:parallel_duration}
\end{subfigure}
\caption{SO posts' required time to receive an accepted answer.}
\label{fig:post_duration_sub}
\vspace{-0.5em}
\end{figure*}

Although we note a high proportion of outliers in Fig.\ref{fig:post_duration}, we have a relatively low median, first, and third quartiles, for all categories compared to the average. Meaning that, although the majority of posts received an accepted answer in a relatively short period of time (in general less than 10 days), a sizable number of posts required a longer time (more than 20 days). Therefore, even within categories, there is quite some discrepancies inside the subcategories (challenges) themselves. 
As can be seen on Fig.\ref{subfig:dl_issues_duration}, SO posts categorized as \textit{Model} have a higher average waiting time to receive an accepted answer than the other subcategories of the \textit{DL Issues} category. Among subcategories under \textit{DRL Issues}, \textit{hyperparameters} includes SO posts with the highest average required time for receiving an accepted answer (Fig.\ref{subfig:drl_issues_duration}). About \textit{DRL libraries/frameworks}, SO posts belonging to \textit{API usage}, \textit{Dependency} and \textit{Installation} subcategories require the longest average time before receiving an accepted answer (Fig.\ref{subfig:drl_lib_duration}). 
Regarding subcategories within \textit{parallel Processing \& multi-threading}, it is notable that the average duration required to receive an accepted answer for SO posts classified under \textit{multi-processing} subcategory 
exceeds one year. 
It should be also taken into account that the small number of SO posts in this subcategory might bias the results, with respect to the fact that small data may not represent the distribution of classes in the population adequately~\cite{bruer2015designing}.

Although our methodology to measure the difficulty level of addressing challenges aligns with prior studies analyzing SO posts~\cite{alshangiti2019developing,decan2019empirical}, 
it is worth noting that some studies have utilized the number of posts within each category as a metric to illustrate the difficulty associated with addressing the related challenges~\cite{bangash2019developers,hamidi2021towards}. 
In the case of considering the number of posts as an indicator of the difficulty level for tackling challenges (the number mentioned along with each category in Fig. \ref{fig:taxonomy}), 
the results would closely mirror the difficulty levels observed for the challenges categorized as \textit{DL issues} and \textit{parallel processing} categories depicted in Fig.\ref{fig:post_duration_sub}. 
For instance, \textit{design problem} is the second most difficult challenge in the \textit{DRL issues} category when we use `number of posts' or `duration to receive an accepted answer' to measure the difficulty level.
Conversely, the difficulty levels for challenges falling under \textit{DRL issues} and \textit{DRL libraries/frameworks} differ by using these two metrics. For example, as illustrated in Fig.\ref{subfig:drl_issues_duration}, \textit{hyperparameters} emerges as the most challenging subcategory within the \textit{DRL issues} category, even though \textit{comprehension} has the highest number of posts. 
This discrepancy may be attributed to the nature of inquiring about \textit{hyperparameters}, which may necessitate various implementations and a longer time to respond compared to other challenges. Furthermore, addressing SO posts categorized as \textit{comprehension} mainly seeks users' background knowledge, which does not necessarily require practical implementation or application execution.

\begin{tcolorbox}[colback=blue!5,colframe=blue!40!black]
\textbf{Finding 2:} 
The proportion of SO posts containing questions related to DRL application development increased significantly during the period (2016 - 2021), reaching a peak between 2019 and 2020. In terms of average time before the reception of an accepted answer, \textit{Parallel processing} is the category that required the longest period of time, in comparison to the other categories. Overall, we observe a high variance in the time to answer of the posts from the different categories of challenges identified in our taxonomy, suggesting that their impact on the development process of DRL applications is not equal. 
\end{tcolorbox}


\subsection*{\textit{RQ2:} How are the identified challenges perceived by DRL practitioners?}

We cross-check the taxonomy generated based on SO posts using a validation survey. \textbf{65} ML practitioners participated in our survey to assess our identified challenges, including 55\% researchers (Master's and Ph.D. students, research assistants, and professors), 27\% ML/SE engineers, 11\% developers, and 7\% data scientists. Among our respondents, 86\% have at least 1 year of DRL development experience and 46\% have more than 3 years of experience. Table~\ref{tbl:validation_report} summarizes the responses of participants for each DRL development challenge contained in the taxonomy.
For each challenge, we provide the percentage of developers who reported having experienced that challenge (based on the answers to the \textit{`yes'} or \textit{`no'} question). Besides, we asked the participants about the severity of each challenge and the effort required to address them. Results show that all challenges presented in our taxonomy were encountered by the survey respondents. Moreover, no additional challenges were proposed by the survey participants through the open-text questions. 
This indicates that our provided taxonomy is representative of challenges faced by developers during the DRL application development. 
It is worth noting that the survey questions incorporate the challenges depicted in the third level of the taxonomy (Fig.\ref{fig:taxonomy}), ensuring the survey's conciseness.

\begin{table}
\caption{Result of the survey of DRL practitioners. 
}
    \centering
    \resizebox{0.95\columnwidth}{!}{
    \begin{tabular}{|c | p{3.5cm}| c c| c c c | c c c |}
        \hline
            \multicolumn{2}{|l|}{\multirow{2}{6em}{
            \textbf{Challenges}
            }} & \multicolumn{2}{ c |}{\textbf{Response}} & \multicolumn{3}{ c |}{\textbf{Severity}} & 
            \multicolumn{3}{ c |}{\textbf{Required effort}} \\
                 \multicolumn{2}{|l|}{} & Yes & No & Minor & Major & Critical & Low & Medium & High \\
             \hline \\[-2ex]
             \multirow{8}{*}{\begin{turn}{90}DRL Issues\end{turn}}& Comprehension & \cellcolor{blue!15} \textbf{48\%} & 52\% & 69\% & 23\% & \cellcolor{blue!15} \textbf{8\%} & 60\% & 23\% & \cellcolor{blue!15} \textbf{17\%} \\[0.2ex]
             & Reward & \cellcolor{blue!15} \textbf{86\%} & 14\% & 32\% & 37\% & \cellcolor{blue!15} \textbf{31\%} & 28\% & 28\% & \cellcolor{blue!15} \textbf{44\%} \\[0.2ex]
             & Action & 49\% & 51\% & 74\% & 11\% & 15\% & 66\% & 20\% & 14\% \\[0.2ex]
             & Environment & \cellcolor{blue!15} \textbf{83\%} & 17\% & 32\% & 34\% & \cellcolor{blue!15}\textbf{34\%} & 31\% & 29\% & \cellcolor{blue!15} \textbf{40\%} \\[0.2ex]
             & Hyperparameters & \cellcolor{blue!15} \textbf{80\%} & 20\% & 43\% & 32\% & \cellcolor{blue!15} \textbf{25\%} & 37\% & 26\% & \cellcolor{blue!15} \textbf{37\%} \\[0.2ex]
             & Design problem & \cellcolor{blue!15} \textbf{75\%} & 25\% & 38\% & 37\% & \cellcolor{blue!15} \textbf{25\%} & 34\% & 31\% & \cellcolor{blue!15} \textbf{35\%} \\[0.2ex]
             & Policy & 62\% & 38\% & 54\% & 37\% & 9\% & 54\% & 32\% & 14\% \\[0.2ex]
             & State/Observation & 68\%  & 32\% & \cellcolor{blue!15} \textbf{52\%} & 32\% &   16\% & 49\% & 37\% & 14\% \\[0.2ex]
             \hline \\[-1.6ex]
             \multirow{6}{*}{\begin{turn}{90}
                \parbox[c]{2.1cm}{\centering
                DRL libraries/ frameworks}
             \end{turn}}& Installation &  \cellcolor{blue!15} \textbf{43\%} & 57\% & 85\% & 12\% & \cellcolor{blue!15} \textbf{3\%} & 78\% & 15\% & \cellcolor{blue!15} \textbf{7\%} \\[0.2ex]
             & Dependency &  54\% & 46\% & 78\% & 14\% &  8\% & 77\% & 18\% & \cellcolor{blue!15} \textbf{5\%} \\[0.2ex]
             & API usage & \cellcolor{blue!15} \textbf{38\%} & 62\% & 86\% & 9\% & \cellcolor{blue!15} \textbf{5\%} & 78\% & 14\% & \cellcolor{blue!15} \textbf{8\%} \\[0.2ex]
             & Documentation & 51\% & 49\% & 69\% & 23\% & 8\% & 71\% & 20\% & 9\% \\[0.2ex]
             & Bugs inside framework& 48\% & 52\% & 75\% & 14\% & 11\% & 70\% & 15\% & 15\% \\[0.2ex]
             & Best fitted library & 51\% & 49\% & 70\% & 25\% & 5\% & 65\% & 25\% & 10\% \\[0.2ex]
             \hline \\[-2ex]
             \multirow{4}{*}{\begin{turn}{90}
             \parbox[c]{1.4cm}{\centering
             Parallel processing
             }
             \end{turn}}& GPU usage & 54\% & 46\% & 72\% & 20\% & 8\% & 69\% & 20\% & 11\% \\[0.2ex]
             & Multi-threading & 45\% & 55\% & 68\% & 24\% & 8\% & 65\% & 18\% & 17\% \\[0.2ex]
             & Multi-processing & 49\% & 51\% & 57\% & 31\% & 12\% & 51\% & 23\% & 26\% \\[0.2ex]
             & Distributed systems & 43\% & 57\% & 58\% & 28\% & 14\% & 55\% & 14\% & \cellcolor{blue!15} \textbf{31\%} \\[0.2ex]
             \hline \\[-2ex]
             \multirow{3}{*}{\begin{turn}{90}
             \parbox[c]{1.1cm}{\centering
             DL Issues
             }
             \end{turn}}& Model & 48\% & 52\% & 71\% & 23\% & 6\% & 66\% & 26\% & 8\% \\[0.4ex]
             & Data prepossessing & 48\% & 52\% & 57\% & 25\% & 18\% & 63\% & 25\% & 12\% \\[0.4ex]
             & DL framework & 46\% & 54\% & 66\% & 25\% & 9\% & 64\% & 27\% & 9\% \\[0.4ex]
            \hline 
    \end{tabular}
    }
        \label{tbl:validation_report}
\end{table}

According to the survey results, the majority of our respondents have been confronted with challenges classified as \textit{DRL issues} ($68.9\%$ average over all subcategories in this category). 
This observation is aligned with the proportion of SO posts categorized as \textit{DRL issues} (see Fig.\ref{fig:taxonomy}).
Conversely, challenges belong to \textit{Parallel Processing \& Multi-threading} category
have been experienced the least with only $45.25\%$ of respondents (which is the lowest proportion among all categories) reporting having faced challenges leveraging parallel processing. 
This finding is reflected by the results of our quantitative analysis of SO posts, which show that only $1.5\%$ of posts contained questions related to the  \textit{Parallel Processing \& Multi-threading} category. 
It should be also taken into consideration that previous research showed that there is a growing trend toward the number of studies on DRL~\cite{panzer2022deep,kiran2021deep}.
Among the challenges identified in our taxonomy, developers who participated in our survey specify \textit{reward} (86\%), \textit{environment} (83\%), \textit{hyperparameters} (80\%), and \textit{design problem} (75\%) as the most common challenges in DRL development. Although only 14.8\% of SO posts contained questions about \textit{API usage}, 38\% of survey respondents identified it as a challenging issue in DRL development. Given that \textit{reward}, \textit{environment}, \textit{hyperparameters}, and \textit{design problem} are fundamental components of an RL application~\cite{lorenz2022reinforcement}, it is expected that survey participants reported them as the most encountered challenges. For instance, defining the environment is known as a crucial step in an RL application development process that affects the convergence of an agent's behavior significantly~\cite{reda2020learning}. 

This however contrasts with the number of SO posts identified as \textit{DRL environment} (i.e., 1.6\% of SO posts). We also note that \textit{comprehension}, the most frequent challenge in terms of the number of SO posts in our taxonomy (28.6\% of SO posts), has been reported by 52\% of survey participants as a non-challenging issue. The explanation for this variance lies in the experience level of the survey respondents. 
Indeed, it is indicated that 84\% of the survey participants have at least 1 year of experience in DRL development 
In other words, experienced practitioners are less likely to seek help for understanding the fundamental DRL concepts because they have already mastered these basics. Moreover, it can be interpreted as the fact that the most challenging steps in the development of DRL applications for DRL practitioners are related to providing an optimized solution for various DRL-related problems, not just addressing a DRL-related problem. Therefore, to fulfill the specific requirements of various DRL developers with different experience levels, it is important to acknowledge that DRL developers have unique needs at different stages of the DRL development journey. Moreover, it should be taken into consideration that the survey was conducted in 2023; nearly a decade after DRL started to become mainstream~\cite{mnih2015human,li2017deep}. 

To enhance the completeness of our provided taxonomy, we scrutinize 254 sampled commits and issues extracted from real-world DRL-related repositories (Subsection~\ref{sec:taxonomy_const}). 
Upon thorough examination of the sampled commits, we did not identify any instances of challenges being mentioned in relation to the DRL. Furthermore, an analysis of sampled closed issues revealed a consistent pattern wherein users primarily seek support on the utilization of DRL applications offered by the repositories.

According to our 13-year analysis of the distribution of DRL-related SO posts, there has been a drop in the number of SO posts categorized as \textit{DRL issues} after its peak in 2019 (Fig.\ref{sub_fig:all_posts_per_year} and Fig.\ref{sub_fig:drl_posts_per_year}). This phenomenon may be attributed to various factors. Initially, it suggests a potential progression in the mastery of fundamental DRL concepts among developers over these years leading to a reduction in challenges encountered and thereby a decrease in the number of DRL-related posts on SO. This can stem from the fact that the growth of DRL popularity in the community results in increased accessibility of DRL tutorial resources including books, tutorials, videos, and papers. These resources aid DRL developers in enhancing their understanding of foundational DRL concepts. Besides, these resources mostly address various DRL problems, including repositories of practical DRL examples that facilitate comprehension of DRL concepts. Moreover, as time has passed, the accumulation of SO posts regarding DRL development has delivered a rich source of DRL development challenges. As a result, many DRL developers can potentially find answers to their questions among the existing SO posts. It should be also taken into consideration that previous research showed that there is a growing trend toward the number of studies on DRL~\cite{rao2018deep,li2017deep}, so the drop in DRL-related SO posts does not imply a drop in the popularity of research on this topic.

\begin{tcolorbox}[colback=blue!5,colframe=blue!40!black]
\textbf{Finding 3:} 
Survey respondents encountered all the challenges included in our taxonomy. They identified \textit{reward}, \textit{design problem}, \textit{environment}, and \textit{hyperparameters} as the most common challenges in the development of DRL applications. However, this observation highlights a notable contrast with the findings from the analysis of SO posts, wherein \textit{comprehension} with 28.6\% of SO posts is the most frequent challenge. The variability observed can be attributed to the fact that DRL practitioners mostly face challenges when they are trying to provide solutions for different DRL problems. 
\end{tcolorbox}


We also ask survey respondents about the severity and needed effort to address the challenges identified in the provided taxonomy (Table \ref{tbl:validation_report}). In general, the majority (exceeding $57\%$) of the survey respondents indicated that the most frequent challenges from their viewpoint 
(i.e., \textit{Reward}, \textit{Environment}, \textit{Design problem} and \textit{Hyperparameters}) are major or critical. 
Moreover, at least 
$63\%$
of the survey participants considered the level of effort required to address these challenges, to be ``Medium" or ``High". 
The majority (more than $52\%$) of the survey participants consider the other identified challenges to be of 
\textit{Minor} severity level, and to require a \textit{Low} level of effort. 
In general, the participants consider that \textit{Installation} and \textit{API usage} challenges require a low level of effort, which might signal that 
DRL libraries/frameworks have good documentation and usability in general~\cite{mojica2023machine}.

We compared the time-to-answer of the posts (from different challenges categories) with the effort reported by our survey participants for the different challenges categories and made the following observations:  

\begin{itemize}
    \item \textit{Hyperparameters}, and \textit{design problems} are the subcategories of \textit{DRL issues} that took longer time before receiving an accepted answer. Survey respondents also reported them as severe and requiring a high effort from ML developers.
    \item The average time required to receive an accepted answer for \textit{State/observation} and \textit{comprehension} SO questions are comparable (even though some SO posts within \textit{comprehension} subcategory took a bit longer to receive an accepted answer than posts belonging to \textit{State/observation}). The survey participants also assessed these two subcategories of challenges (i.e., \textit{comprehension} and \textit{State/observation}) as easy to resolve in general. 
    \item \textit{API usage} and \textit{Dependency} challenges are the groups of challenges that took the longest time before their questions received an answer. This result is in contrast with the survey participants' estimation of the effort required to fix them.  
\end{itemize}
The severity and required effort reported by the survey participants for each challenge are strongly correlated (using kendall’s tau~\cite{gibbons1993nonparametric}) in a positive direction~\cite{frost2019introduction} and statistically significant ($P-value < 0.05$). Hence, more severe challenges necessitate more effort from developers.


\begin{tcolorbox}[colback=blue!5,colframe=blue!40!black]
\textbf{Finding 4:}  
Survey respondents highlighted \textit{Reward, Environment, Hyperparameters}, and \textit{Design problem} as the most severe and intricate-to-address challenges in DRL development. 
\end{tcolorbox}





\subsection*{\textit{RQ3:}  Are DRL application development challenges language- and/or framework-specific? 
}

\begin{figure}
    \centering
    \includegraphics[width=0.65\textwidth]{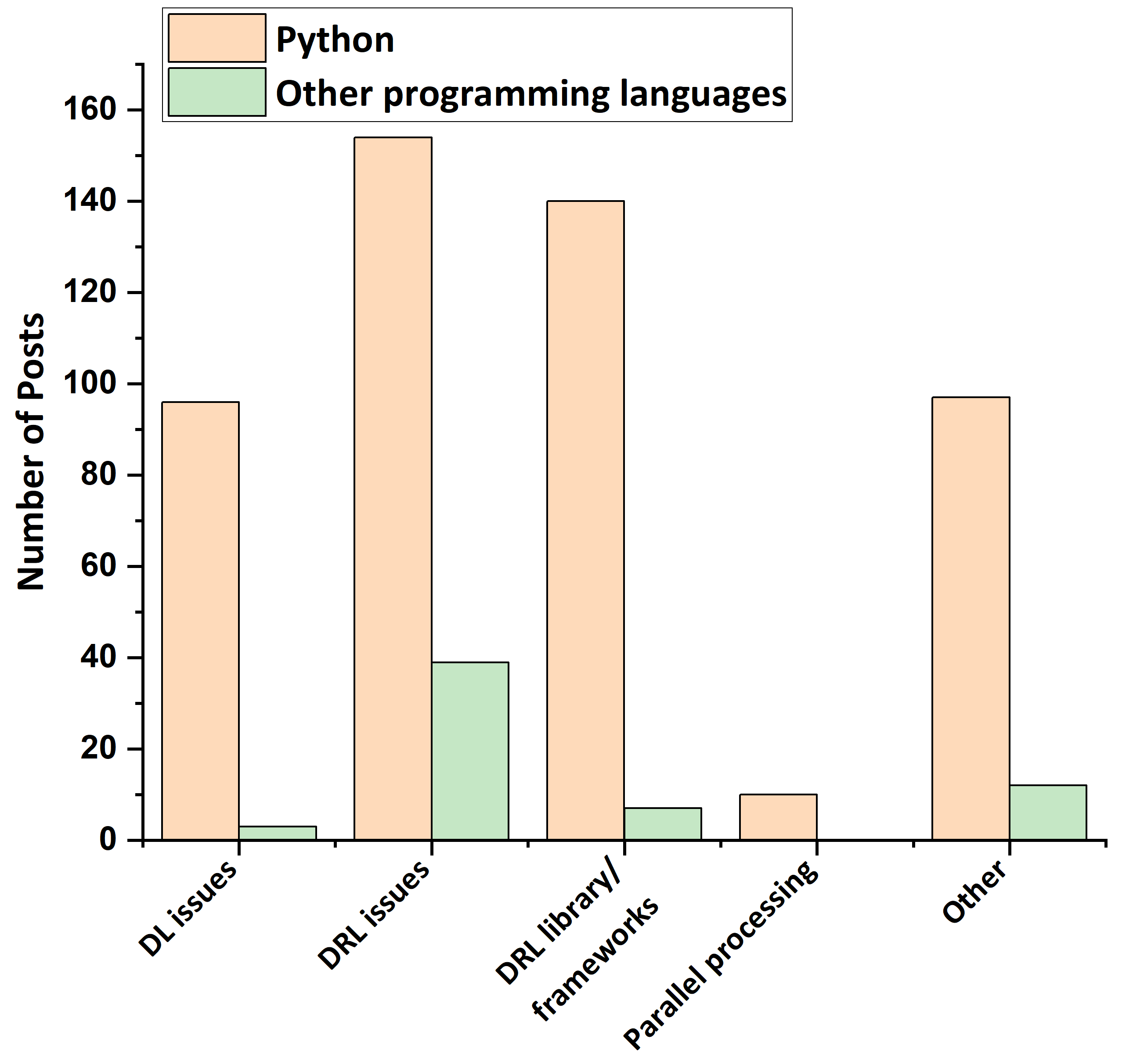}
    \caption{Programming languages mentioned in the posts belong to various categories of challenges. 
    }
    \label{fig:categories-programming-langs}
\end{figure}

We extract information about the programming languages used to develop DRL applications from the collected SO posts. It should be mentioned that the posts have been collected 
without any distinction on the programming language and frameworks used. 
Fig.\ref{fig:categories-programming-langs} presents the proportion of posts using \textit{Python} programming language for the different identified categories of challenges. 
As can be seen, \textit{Python} is by far the dominant programming language for all categories of challenges. However, the proportion of posts mentioning other programming languages and containing \textit{DRL issues} is non-negligible (i.e., 20.2\%).  
This high ratio is mostly attributable to \textit{Java} (5.2\% of all posts), \textit{C++} (4.7\% of all posts), and \textit{R} (4.7\% of all posts) programming languages. 
It is also noteworthy that investigating 
the relationship between used programming languages and the challenges (subcategories) within each category reveals that \textit{Python} stands out as the predominant
programming language across all DRL development challenges. 
Based on these results, we conclude that there is no relationship between DRL development challenges and used programming languages.
This finding is in accordance with prior research~\cite{morovati2023bugs,humbatova2020taxonomy} which reported that \textit{Python} is the most popular programming language for ML-enabled applications. 
These results about programming languages used in DRL applications development are also supported by our validation survey where all participants mentioned \textit{Python} as the programming language they use for developing DRL applications. Besides, 20\% of participants reported \textit{C/C++}, and 12\% mentioned other programming languages in addition to \textit{Python} (e.g., \textit{C\#} 4\%, and \textit{Java} 3\%).

\begin{figure*}[!h]
\centering
\begin{subfigure}[t]{0.45\textwidth}
    \includegraphics[width=0.9\textwidth]{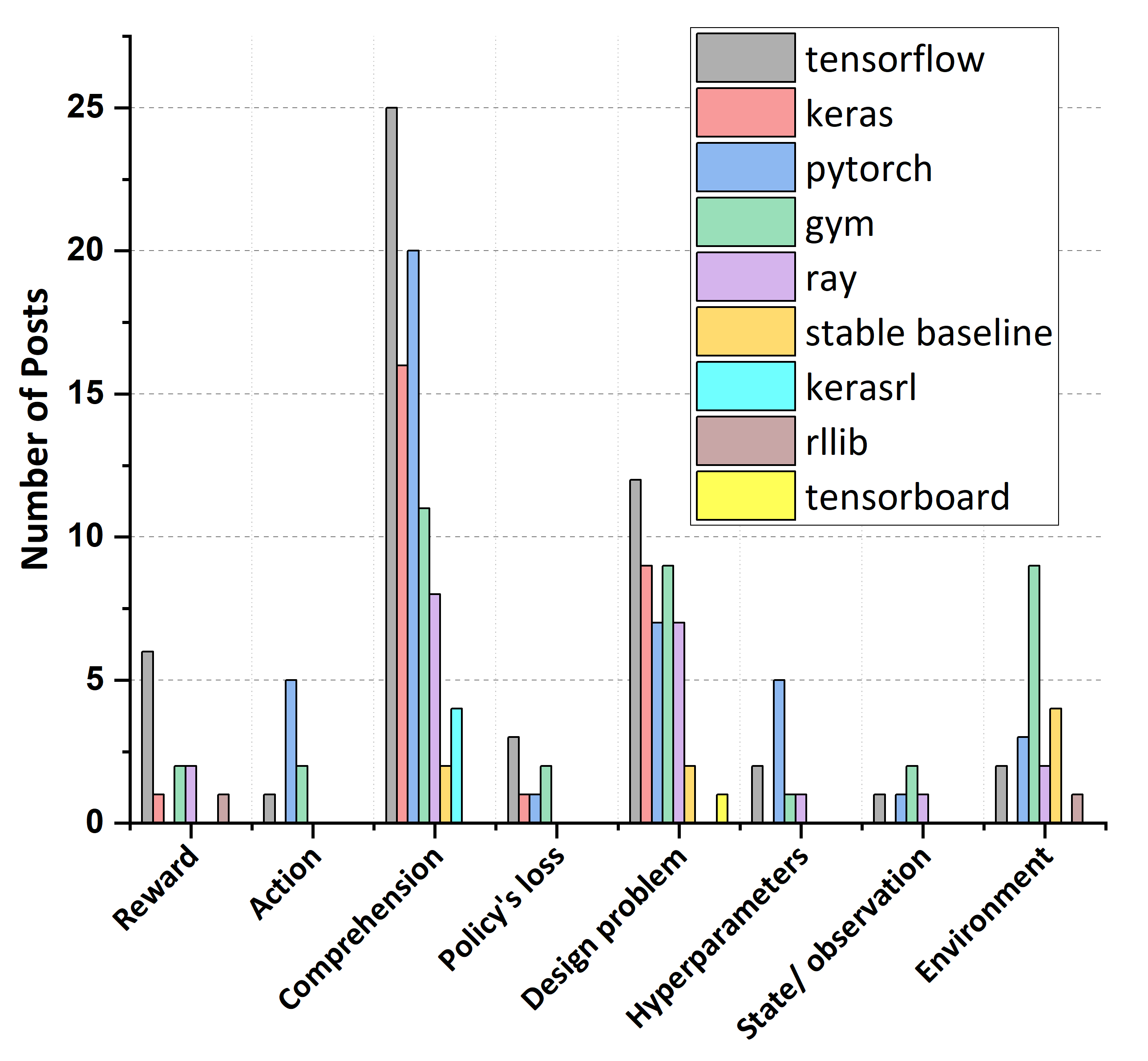}
    \caption{DRL issues}
    \label{subfig:drl-issue-frameworks}
\end{subfigure}\hspace{\fill} 
\begin{subfigure}[t]{0.45\textwidth}
    \includegraphics[width=0.9\linewidth]{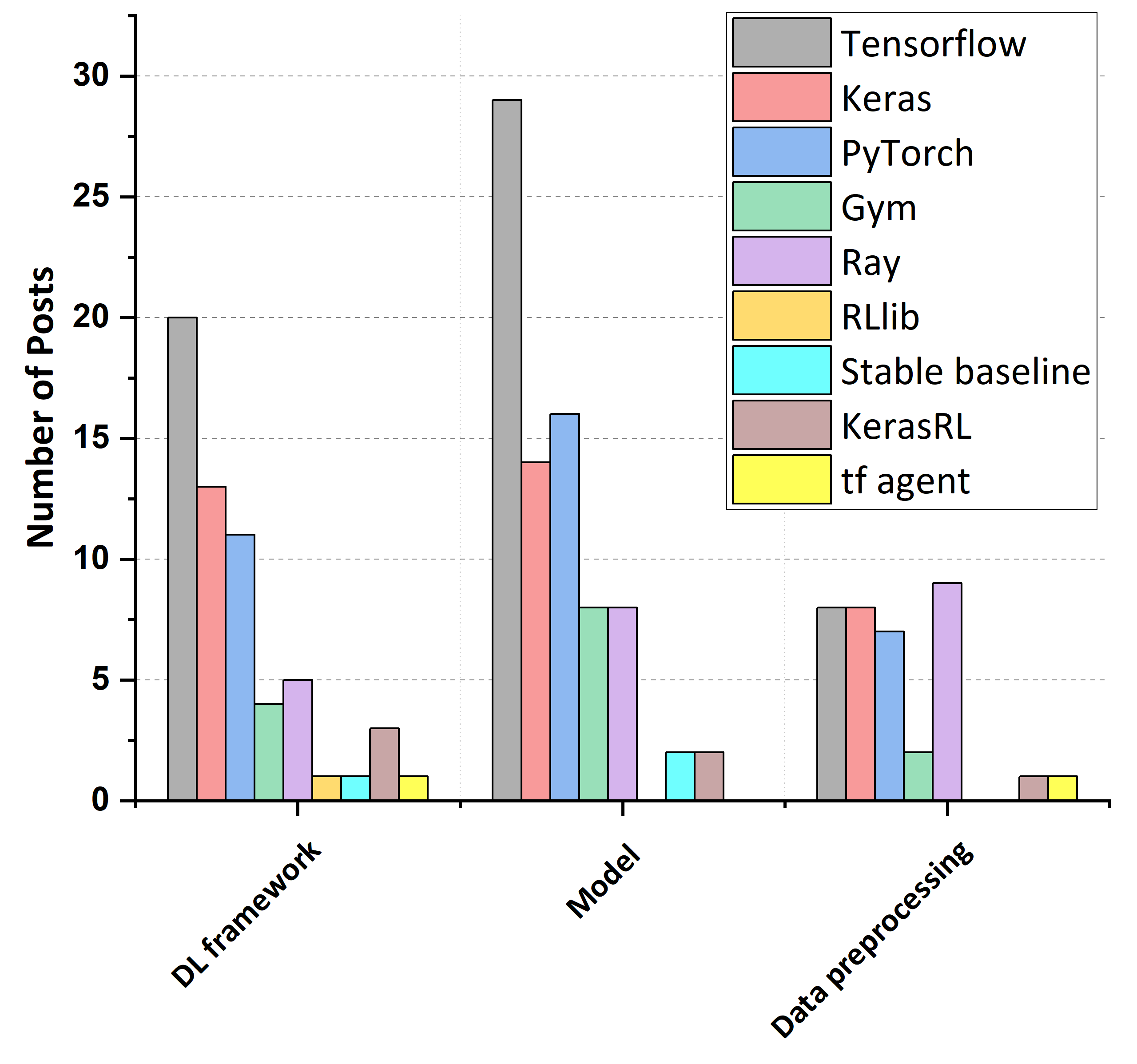}
    \caption{DL issues}
    \label{subfig:dl_issue_frameworks}
\end{subfigure}

\bigskip 
\begin{subfigure}[t]{0.45\textwidth}
    \includegraphics[width=0.9\linewidth]{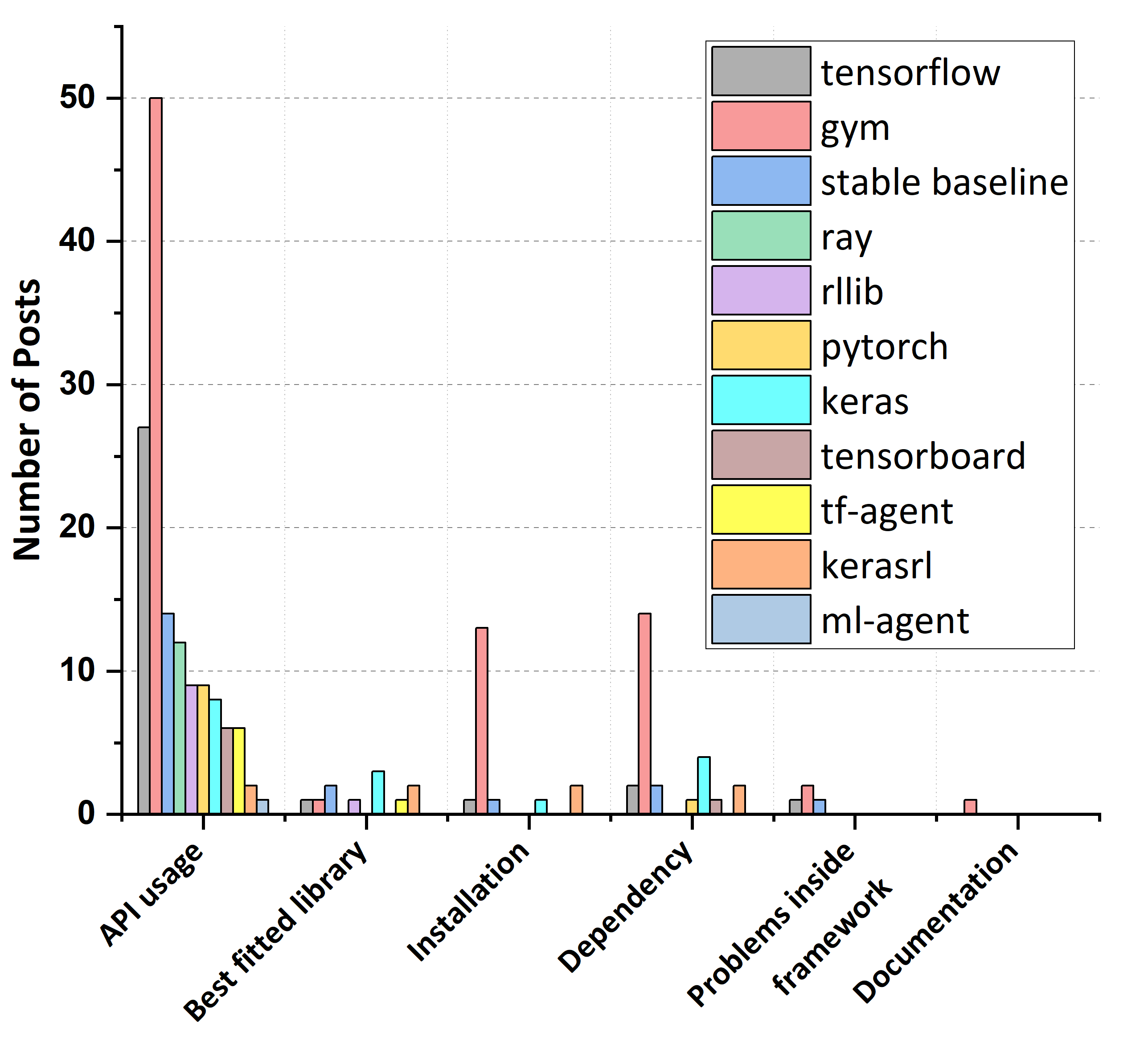}
    \caption{DRL libraries/frameworks}
    \label{subfig:drl_Library_frameworks}
\end{subfigure}\hspace{\fill} 
\begin{subfigure}[t]{0.45\textwidth}
        \includegraphics[width=0.9\linewidth]{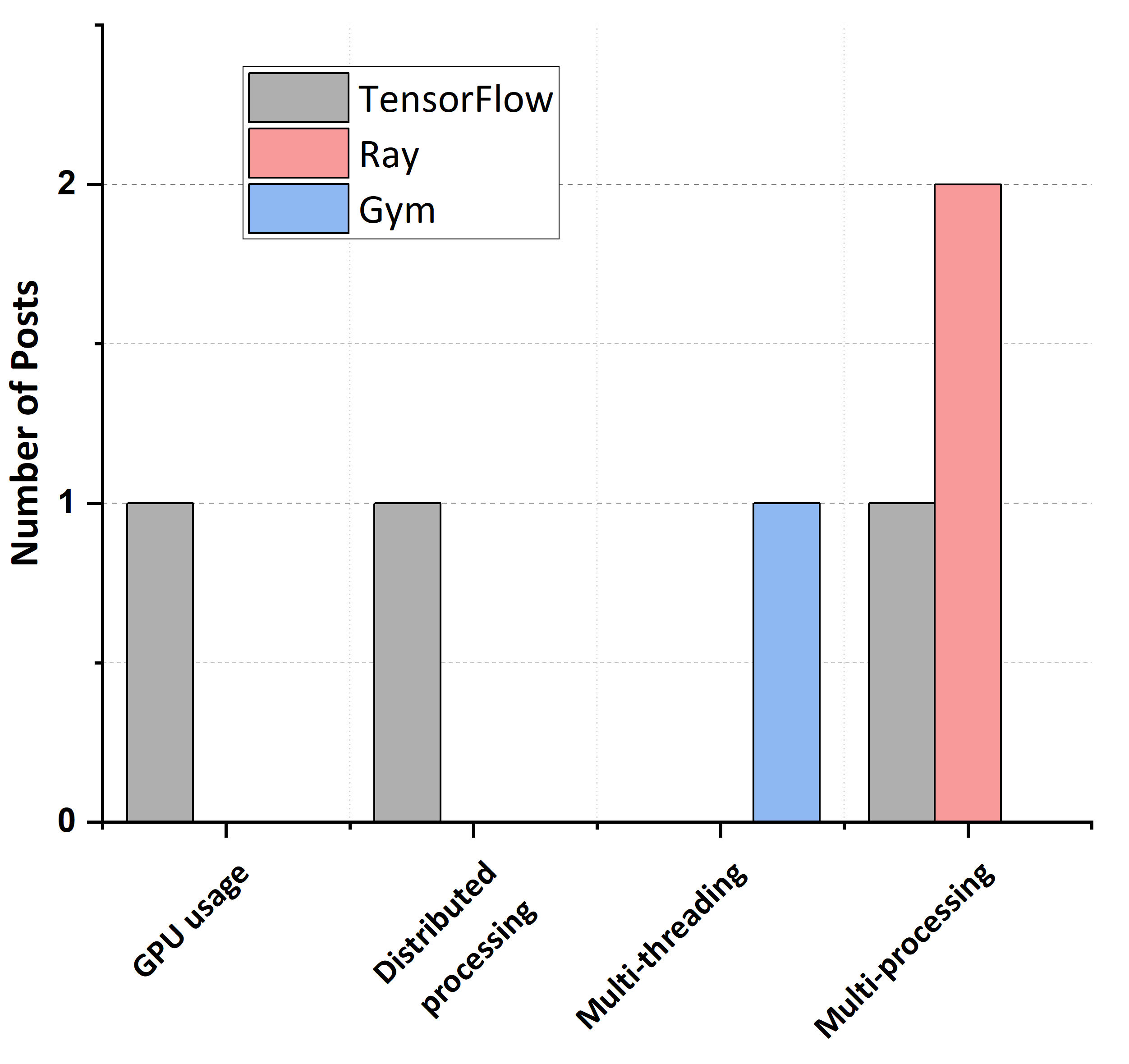}
        \caption{Parallel processing \& multi-threading}
        \label{subfig:parallel_framework}
\end{subfigure}
\caption{Common libraries/frameworks used for developing DRL applications.}
\label{fig:categories-framework}
\vspace{-0.5em}
\end{figure*}

We also examine the mentioned frameworks and libraries in posts related to different challenge categories.
Fig.\ref{subfig:drl-issue-frameworks} shows the number of times that each library/framework has been mentioned in posts belonging to various subcategories of the \textit{DRL issues} category. With an exception for the \textit{environment} and \textit{state/observation} subcategories, \textit{Keras}, \textit{Tensorflow}, and \textit{PyTorch} are the most used libraries/frameworks by SO users, and they are the most popular libraries/frameworks for developing ML and DL\cite{morovati2024bug}. 
Considering that there are several libraries/frameworks specifically designed to ease DRL application development (e.g. \textit{KerasRL}, \textit{RLlib}, etc), Fig.\ref{subfig:drl-issue-frameworks} exposes that SO users usually prefer to use popular ML libraries/frameworks which can be leveraged to implement DRL applications as a subdomain of ML. On the other hand, \textit{gym}~\cite{brockman2016openai} is the most popular library/framework in the \textit{environment} subcategory which is reasonable as it is the most popular library for implementing various RL environments and provides a standard benchmark containing a large number of well-known RL environments~\cite{panerati2021learning}. 
This observation can be related to the inherent nature of SO posts, which may not necessarily provide details about the libraries/frameworks employed by the user posting questions. It is noteworthy that 46\% of 885 examined SO posts lack any reference to the used DRL-related libraries/frameworks. As an example, post \href{https://stackoverflow.com/questions/56312962}{\#56312962} serves as an illustrative case where no information has been mentioned regarding the utilized libraries-frameworks to implement the DRL application. 

Fig.\ref{subfig:dl_issue_frameworks} presents the libraries/frameworks mentioned in the posts classified as \textit{DL issues}. Results show that \textit{TensorFlow}, \textit{Keras}, and \textit{PyTorch} are the most popular libraries/frameworks in the \textit{DL framework} and \textit{model} subcategories. Given that the challenges within the \textit{DL issues} category pertain to the DL parts of DRL applications, it is not surprising to see \textit{TensorFlow}, \textit{Keras}, and \textit{PyTorch} are frequently mentioned since they are the most used libraries in the development of DL-enabled applications~\cite{morovati2023bugs,humbatova2020taxonomy}. 
Moreover, the generality of \textit{Ray} in the \textit{Data preprocessing} subcategory, compared to other libraries/frameworks can be attributed to the fact that \textit{Ray} encompasses not only DRL-related libraries (e.g., \textit{RLlib}~\cite{liang2018rllib}) but also various other libraries for a wide range of ML-related tasks at the same time, including scalable datasets, model training, and hyperparameter tuning~\cite{pumperla2023learning} which may ease developing DRL applications. 

About the most frequently referenced libraries/frameworks in the \textit{DRL libraries/frameworks} category, as illustrated in Fig.\ref{subfig:drl_Library_frameworks}, \textit{gym} received the largest number of questions, especially on the topics of \textit{API usage}, \textit{installation}, and \textit{dependency} challenges. Although \textit{gym} is the most popular library for implementing RL/DRL environments~\cite{panerati2021learning}, comparing Fig.\ref{subfig:drl-issue-frameworks} and \ref{subfig:dl_issue_frameworks} may indicate that \textit{gym} has less matured documentation and tutorials, in comparison to \textit{Keras}, \textit{Tensorflow}, and \textit{PyTorch}. It is also worth mentioning that \textit{Keras}, \textit{Tensorflow}, and \textit{PyTorch} are general-purpose ML-related libraries/frameworks which are more developed, compared to \textit{gym} which is implemented specifically for RL development~\cite{open_ai_gym}. 

Regarding parallelization and multi-threading, as can be seen in Fig.\ref{subfig:parallel_framework}, the majority of issues are reported against \textit{TensorFlow}, particularly regarding \textit{GPU usage} and \textit{distributed processing}. It can be related to the fact that \textit{Tensorflow} is considered the most popular ML framework~\cite{openja2022studying}. On the other hand, most of the \textit{multi-processing} challenges relate to the \textit{ray} framework, which could be attributed to the fact that \textit{ray} supports multiprocessing and RL at the same time~\cite{moritz2018ray}. 


The respondents of our survey corroborated these findings; with 86\% of them reporting \textit{PyTorch} as their preferred framework, followed by \textit{TensorFlow} (50\%), and \textit{Keras} (39\%). Some participants also mentioned \textit{KerasRL} (6\%) and \textit{JAX} (6\%).
We note that \textit{PyTorch} was cited more than \textit{Tensorflow} and \textit{Keras} by participants of our validation survey compared to SO posts. This can be explained by the fact that SO posts do not necessarily include information about the frameworks used by users asking questions.
It is also worth mentioning that we collected SO posts over a period of 13 years (from 2009 to 2022), while \textit{PyTorch} was introduced only in 2016 (in comparison to \textit{Tensorflow} and \textit{Keras} released in 2015).

\begin{tcolorbox}[colback=blue!5,colframe=blue!40!black]
\textbf{Finding 5:} 
\textit{Python} is by far the dominant programming language used for DRL development. 
While \textit{TensorFlow}, \textit{Keras}, and \textit{PyTorch} are frequently mentioned in posts reporting challenges faced by DRL developers, 
\textit{gym} is the most challenging libraries to install (see \textit{installation}, and \textit{dependency} challenges) and use (see 
\textit{API usage} challenges). Our results also show that \textit{ray} is the most challenging library when dealing with \textit{multi-processing} and \textit{data preprocessing} operations. 
\end{tcolorbox}

\section{Discussion}
\label{sec:discussion}
Based on the findings of our study, in this section, we discuss the state of DRL application development and highlight some research avenues for researchers and practitioners. 

Through this study, we gained 
a thorough understanding of frequently asked questions regarding DRL development to enable the community to explore potential approaches for mitigating these challenges, minimizing errors, and enhancing the reliability of DRL applications. Based on our provided taxonomy, one can see that some challenges faced in the development of DRL applications are common to all types of DL applications. For example, managing dependencies when using DL libraries/frameworks is a prevalent challenge in DL applications. However, dependency management can be more complex in DRL, because of the need for synchronization among a larger number of libraries in the development of DRL applications (e.g., aligning the Python version with the DRL libraries/frameworks and the library that manages RL environment). Similarly to what was suggested by Huang et al.~\cite{huang2022demystifying} to tackle dependency management challenges in DL applications, DRL researchers can provide a dependency knowledge graph 
for DRL libraries/frameworks to mitigate this challenge. 


Regarding the provided taxonomy, it is worth mentioning that all of the categories and subcategories of challenges directly relate to the DRL application development. However, some of the challenges may be observed in other ML-related applications. For instance, all of the challenges belonging to \textit{DRL libraries/frameworks} (e.g. \textit{API usage}, \textit{Installation}, \textit{Dependency}, etc.) have been faced by all developers who use ML/DL libraries/frameworks. But it should be also taken into consideration that all of the investigated SO posts in this study have been achieved after a comprehensive filtering process making sure all of the extracted SO posts are about challenges in DRL application development. On the other hand, \textit{Reward}, \textit{Environment}, \textit{Action}, \textit{State/Observation}, and \textit{Policy} challenges are Specific to DRL applications. It should be also taken into consideration that what makes this taxonomy valuable in the DRL community is the fact that their frequency, importance, and severity would be different in DRL application development compared to other ML-related applications. For example, 14.8\% of challenges in DRL application development are related to \textit{API usage}, whereas it is only 5.3\% in DL-related applications~\cite{humbatova2020taxonomy}.

Finding 2 revealed that $27.3\%$ of DRL development challenges categorized as \textit{comprehension} are related to the lack of sound understanding of basic DRL concepts. 
In other words, $58.4\%$ of posts belonging to \textit{DRL issues} category (DRL-specific category) are about \textit{comprehension} challenge. This finding highlights the need for documentation and tutorials to help 
DRL developers 
who are not experts in DRL, in the development of DRL applications. 
A roadmap for the development of DRL applications would also help developers navigate through the implementation of DRL applications with fewer misunderstandings of DRL concepts. 
The need for such material is emphasized 
by a post\footnote{https://stackoverflow.com/questions/37973108} asking questions about the difference between RL and DRL. By providing a roadmap that systematically expands DRL developers' understanding, developers will be supported in overcoming the most common challenge in DRL application development. 
An illustration of such guidance is the work conducted by Garg et al.~\cite{garg2019roadmap} on creating a roadmap for DL development. The need for good documentation and guidance is also emphasized by the survey participants who mentioned that \textit{`although there are a number of tutorials to start working on DRL, a few issues are shared between many of them'}. The participants also noted that many of the DRL tutorial documents cover only a specific domain of DRL. Participants also lamented the poor usability of 
DRL-related tutorials, claiming that they often contain 
a lot of unnecessary materials. 

Leveraging our Findings 1 and 2, researchers can develop 
debugging tools to help developers identify the issues early on during DRL application development. 
Debugging tools can significantly reduce DRL development and maintenance costs.  
For instance, considering the limited documentation available for most of the DRL libraries/frameworks in comparison with DL libraries for example (e.g., \textit{TensorFlow}), a helpful approach would be proposing techniques and tools to assist DRL developers when using different DRL APIs. 
This could help mitigate DRL API issues. 
An example of such techniques focusing on the challenges of software API usage, is the work by Xie et al.~\cite{xie2022docter} which proposes an approach to automatically extract the API parameter constraints of DL libraries/frameworks.

Finding 2 of this study regarding challenges associated with installation and dependency management of DRL libraries/frameworks is aligned with previous studies on dependency management in software development~\cite{cao2022towards} in general, as well as in DL applications~\cite{han2020empirical}. 
Considering the complex nature of DRL application development (due to the communication of several libraries), challenges regarding libraries/frameworks installation and their dependency management become more intricate, compared to other types of DL applications. 
This highlights the need for tools 
(e.g., package manager) to support dependency management. 
As an example, researchers can provide a tool (such as Maven\footnote{https://maven.apache.org/} for Java) that automates the identification and installation of DRL libraries/frameworks that are best suited for a specific Operating System (OS) and a specific version of \textit{Python}. Additionally, such tools can assist DRL developers in synchronizing the installed DRL-related libraries/frameworks when they need to update some of them.

Our results in Finding 2 also stress the need for supporting tools and documentation, for 
parallel and distributed DRL application developments. Questions related to this topic took a long time before receiving an accepted answer on Stack Overflow. 
DRL experts could consider developing pre-configured packages to support 
parallel/distributed DRL application developments. 
As an example of the same task in the ML-related development, 
Openja et al.~\cite{openja2022studying} examined ML application deployment practices on Docker and reported that a significant number of ML developers use Docker to manage dependencies, environment, and the execution of ML applications.




\section{Related Works}
\label{sec:related_work}
We now report and discuss the related literature.  
\subsection{SO Posts Analysis}
Beyer et al.~\cite{beyer2020kind} investigated the automatic classification of 
SO posts. They manually labeled 1000 posts, and identified 7 categories of questions: 1) API changes, 2) API usage, 3) Conceptual, 4) Discrepancy, 5) Learning, 6) Errors, and 7) Review. Leveraging the labeled dataset, they developed 
two approaches for the automatic classification of SO posts. In the first approach, they used the labeled dataset to extract some regular expression patterns and used these patterns to predict the category of other posts; achieving a performance of $0.91$ for both precision and recall. In the second approach, they trained Random Forest and Support Vector Machine (SVM) classifiers using the labeled dataset. The best results were obtained using the 
Random Forest classifier, i.e., a precision of 0.88 and a recall of 0.87.

Alshangiti et al.~\cite{alshangiti2019developing} investigated SO posts related to ML development. They used a tag-based snowball sampling approach to extract SO posts related to ML, starting with the `machine-learning' tag. Their results revealed that a higher number of ML-related posts remain without any accepted answer (61\%), in comparison with general domain questions (48\%). 
They also reported that ML-related questions need 10 times longer to receive an answer, compared to general domain questions. Next, they compared the ratio of expert users in ML and web development (the most popular domain of programming in SO), showing that the number of ML experts is significantly less than that in web development. 
Afterward, they reviewed the most challenging ML development phases revealing that data preprocessing and manipulation, and model deployment and environment setup are the two most error-prone phases. 

Bangash et al.~\cite{bangash2019developers} conducted an empirical study of SO posts related to ML. They used the `machine\_learning' tag to extract 28,010 posts published between 2008 and 2018. Next, they used Latent Dirichlet allocation (LDA)~\cite{jelodar2019latent} to categorize the extracted posts into 44 detailed topics. Then, they showed that \textit{code errors}, \textit{Algorithms}, and \textit{Labeling} are the most discussed topics in ML. They then combined the 44 topics identified using LDA into 4 main groups including frameworks, implementation, sub-domain (RE), and algorithms. They reported that nearly $51\%$ of all ML-related SO posts belong to the implementation group. Afterward, two of the researchers manually examined 230 sampled posts and reported that most of the questions stem from the fact that ML novice developers try to use ML in their software systems. They extracted information about the number of questions with accepted answers and concluded that ML-related questions are harder to answer than general domain SO questions. They also observed that only $65.6\%$ of ML-related SO posts have appropriate tags; which might suggest that many users are not knowledgeable enough to assign the proper ML-related tags to their posts.

Hamidi et al.~\cite{hamidi2021towards} examined the challenges that developers may face in the development of ML systems, based on their discussion in SO. They studied $43,950$ ML-related SO posts submitted between 2008 and 2020. First, they showed that \textit{Python} is the most popular programming language for ML development, and C\# and C/C++ are the least popular programming languages for ML development. Then, they reported that model building and model evaluation are the two most challenging steps in ML development while model monitoring is the least questioned phase. They also report that questions regarding model requirements, data collection/processing, and model-building steps receive less accepted answers than others. This may stem from the fact that questions about these steps are more difficult to answer or the lack of active knowledgeable developers on SO to answer questions related to these steps. 

Although these previous works investigated SO questions related to ML/DL application development, to the best of our knowledge, none of them examined the challenges of DRL development specifically. 

\subsection{RL and DRL Quality Assurance}
In this section, we report on studies about the quality assurance of RL and DRL applications.  

Zhang et al.~\cite{zhang2021autotrainer}  proposed strategies to help DL and DRL developers detect and resolve quality issues in their applications. 
Nikanjam et al.~\cite{nikanjam2021automatic} proposed a methodology for automatically detecting faults in DL applications, using graph transformations. 
Schoop et al.~\cite{schoop2021umlaut} also proposed a strategy for detecting common issues during the 
development of DL models. These studies primarily targeted issues occurring in the training program of DL models. The issues considered in these aforementioned study 
fall within the \textit{`DL issues'} category of our provided taxonomy, which constitutes only a small portion of DRL issues. 

Nikanjam et al.~\cite{nikanjam2022faults} also investigated challenges categorized as \textit{DRL issues} in our proposed taxonomy. They examined questions/discussions about four popular DRL frameworks (including \textit{gym}~\cite{brockman2016openai}, \textit{Tensorforce}~\cite{tensorforce}, \textit{Dopamine}~\cite{castro18dopamine}, and \textit{Keras-rl}~\cite{plappert2016kerasrl} on GitHub and SO) and extracted 329 SO posts about DRL. They categorized these posts into six groups: basic concepts, without acknowledgment, implementation issues, answered by the owner, relative questions, and others. They reported that `without acknowledgment' and `implementation' questions are the most common DRL-related questions in SO, accounting for 32\% and 27\%, respectively. They also showed that in 2\% of their studied SO posts, the answer has been posted by the questioner. They report that DRL-related SO posts take an average and median time of 2.07 days and 13 hours, respectively, before receiving an accepted answer. 
This period is longer than the time taken by DL-related SO posts to receive an accepted answer; which is 5 hours on average. This finding implies that DRL-related questions might be more difficult to answer than DL-related questions. 
Nikanjam et al.~\cite{nikanjam2022faults} also proposed a taxonomy of faults in DRL models. with 11 different issue types. Our study differs from what Nikanjam et al.~\cite{nikanjam2022faults} carried out in several aspects. Firstly, they concentrated on four specific libraries/frameworks designed for DRL development, ignoring other SO posts containing source codes using other Python-based libraries/frameworks or not mentioning any script. 
For instance, a number of SO posts inquire about DRL concepts  (e.g. \#\href{https://stackoverflow.com/questions/52838439}{52838439}) without mentioning any scripts, a scope not covered by Nikanjam et al.~\cite{nikanjam2022faults}. Moreover, their research was just on the DRL model, whereas our study delves into the challenges developers may face throughout the development of entire DRL applications, without any limitation to a specific section of DRL applications. Last but not least, Nikanjam et al.~\cite{nikanjam2022faults} examined SO posts reporting program faults during the development of DRL applications. 
It is imperative to distinguish between challenges and program faults and note that challenges do not necessarily equate to program faults. A fault denotes a defect or error leading to a discrepancy between the expected and achieved results or observed behavior~\cite{morovati2023bugs}. Software development challenges, in contrast, encompass any difficulty or complexity encountered in completing a development task. These challenges may arise from various factors, including technical complexities, resource constraints, or lack of expertise. 
Nikanjam et al.~\cite{nikanjam2022faults} also investigated challenges categorized as \textit{`DRL issues'} in our proposed taxonomy.
They examined questions/discussions about four popular DRL frameworks (including \textit{gym}~\cite{brockman2016openai}, \textit{Tensorforce}~\cite{tensorforce}, \textit{Dopamine}~\cite{castro18dopamine}, and \textit{Keras-rl}~\cite{plappert2016kerasrl} on GitHub and SO) and extracted 329 SO posts about DRL. 
Since Nikanjam et al.~\cite{nikanjam2022faults} focused on software faults, they excluded 305 SO posts from their dataset, generating their taxonomy based on 24 such posts that included faults explicitly. Notably, all SO posts in their dataset are included in our study. Furthermore, our research incorporates SO posts that do not pertain to program faults but instead pose queries related to the comprehension of DRL concepts. For example, the comprehension subcategory includes 253 SO posts, none of which reports program faults.
They categorized these posts into six groups: basic concepts, without acknowledgment, implementation issues, answered by the owner, relative questions, and others. They reported that `without acknowledgment' and `implementation' questions are the most common DRL-related questions in SO, accounting for $32\%$ and $27\%$, respectively. They also showed that in $2\%$ of their studied SO posts, the answer has been posted by the questioner. 
They report that DRL-related SO posts take an average and median time of $2.07$ days and $13$ hours, respectively, before receiving an accepted answer. This time period is longer than the time taken by DL-related SO posts to receive an accepted answer; which is $5$ hours on average. This finding implies that DRL-related questions might be more difficult to answer than DL-related questions. 
Nikanjam et al.~\cite{nikanjam2022faults} also proposed a taxonomy of faults in DRL models. 
The most significant difference between their study and ours is their focus on the DRL model only and the fact that they only collected data about faults mentioned in SO; disregarding all the other types of questions. 

Yahmed et al.~\cite{yahmed2023deploying} conducted a study on the challenges of deploying DRL systems, based on the questions that developers ask on SO. In the first step, they extracted 357 SO posts related to the deployment of DRL systems.  Next, they categorized collected SO posts into 4 categories with respect to their deployment platform, including `server/cloud', `mobile/embedded system', `browser', and `game engine'. Their results showed that the number of SO posts regarding deployment has grown over the 7 studied years. They manually examined the extracted SO posts and identified 31 challenges related to DRL deployment. They grouped these challenges into 11 categories (and proposed a taxonomy): general questions, deployment infrastructure, data preprocessing, RL environment, communication, agent load/save, performance, environment rendering, agent export, request handling, and continuous learning. The proposed taxonomy has been evaluated via a survey with practitioners. Their results show that DRL developers struggle the most with deployment infrastructures and RL environments. Their results also show that communication-related challenges (procedure, connection loss, configuration of remote setting, and model convergence) are the most difficult challenges to address, in terms of the time to an accepted answer.  The main difference between the work of Yahmed et al. and our study is the focus of the study. Yahmed et al. examined DRL deployment challenges while we examined challenges faced by DRL developers during the application development phase, i.e., prior to deployment. In other words, they ignored all DRL development steps before deployment and focused only on the deployment phase of the DRL applications, which occurs after complete implementation of them. At the opposite extreme, we investigated the SO posts regarding the whole pipeline of DRL applications development. Besides, comparing the number of studied SO posts in these two studies (357 vs 927 SO posts) demonstrates that Yahmed et al filtered out the whole dataset of SO posts on DRL development to achieve posts specifically talking about the deployment step of the DRL application. 

\section{Threats to Validity}
\label{sec:validity}
We now discuss threats to the validity of our study.\\
\textit{Construction validity.} Our methodology and labeling process can be a potential validity threat. We have thoroughly described our process and the tags used to collect the posts. As no previous taxonomy on this subject exists, we used an open coding approach with multiple rounds and cross-checking to ensure continuous improvement and consistency of the labeling. We further validate our results via a survey with 65 DRL practitioners.

\textit{Internal validity.} As users do not necessarily provide suitable tags for their questions, our search might have missed some DRL challenges. For example, post \href{https://stackoverflow.com/questions/37973108}{\textit{37973108}} is related to DRL, but it does not have any specific tag mentioning DRL. Nonetheless, tag usage was necessary for a consistent methodology and we believe that the number of posts gathered and analyzed (927) is sizable enough to provide a good representation of the challenges faced by DRL developers. 
Moreover, we used a snowballing approach to expand our basic set of DRL-related tags used to extract DRL-related SO posts, similar to previous studies~\cite{ayman2019smart}. In addition to extracting posts based on the DRL-related tags, we used a set of keywords to extract DRL-related SO posts without any DRL-related tags to address this issue, as other researchers followed a similar methodology~\cite{peruma2022refactor}. 
Another source of threat to the validity of this study arises from the potential overlap between users posing questions in SO posts and the developers associated with the DRL-related GitHub repositories that are used for our survey. To address this concern, we provided a detailed description for each category and subcategory of the taxonomy within the survey, intentionally avoiding any specific information references to any SO post as an example of challenges. Despite these precautions, the prospect of overlap between these two groups of DRL developers remains a possibility.
Although utilizing the duration to receive an accepted answer has been employed in several studies~\cite{alshangiti2019developing,decan2019empirical} as a metric to gauge the difficulty level of SO posts, it may present a potential threat to the internal validity of this research. Reproducibility of issues in ML-enabled systems poses significant challenges and sometimes needs more time compared to traditional software systems~\cite{shah2024towards}. In other words, addressing certain questions in ML-enabled systems (such as those related to parallel processing) requires specific configurations or environments, so the question may be quite simple, but it requires a long time to receive an accepted answer. To address this concern, we also considered the number of posts, which has been used as an indicator to measure the difficulty level of SO posts in a few studies~\cite{bangash2019developers,hamidi2021towards}. However, as discussed in Section \ref{sec:discussion}, this metric yielded similar results to the methodology we employed (i.e., the time to receive an accepted answer).


\textit{External validity.} While there might exist other DRL challenges that practitioners are facing, we conducted our study using SO which is the largest technical Q\&A platform in the software development community. Moreover,
all the challenges identified in our provided taxonomy have been validated by our survey participants. Also, respondents of the survey did not report any other challenge that is not included in our provided taxonomy; which is a good result, with respect to completeness.
Another external factor that could pose a threat to the validity of our results is the possibility that users who raised SO posts may mostly fall into the category of less experienced users. To mitigate this concern, we examined the top 100 DRL-related GitHub repositories to extract the challenges that GitHub developers mention in their development process. However, it appears that experienced developers generally refrain from detailing their challenges in GitHub commit messages or issues. It is noteworthy that our provided taxonomy may represent high-level categories of challenges, covering almost all aspects of DRL application development. Therefore, refining the taxonomy to offer more detailed categorization could be explored as a potential avenue for future research in this study. 

\textit{Reliability validity.} We described our methodology in detail and provided a replication package~\cite{replication_package} to allow others to replicate our results and expand our study. 

\section{Conclusion and Future Works}
\label{sec:conclusion}
In this study, we conducted a large-scale empirical study of 927 DRL-related posts extracted from SO. We examined all posts manually to identify the challenges that developers face when developing DRL applications. We found that \textit{Python} is by far the most popular programming language and \textit{TensorFlow}, \textit{Keras}, \textit{PyTorch}, and
\textit{OpenAI Gym} are the most frequently used libraries/frameworks for developing DRL applications. We categorized DRL development challenges into five groups including \textit{DRL issues}, \textit{parallel processing \& multi-threading}, \textit{DRL libraries/frameworks}, \textit{DL issues}, and \textit{general programming}. An analysis of the received response by the investigated SO posts shows that DRL comprehension, DRL libraries/frameworks API usage, and designing a problem using DRL algorithms are the most challenging parts of DRL applications' development. Furthermore, parallel processing/multi-threading and DRL libraries/frameworks challenges required a longer time to receive an accepted answer. We proposed a taxonomy of challenges and validated it using a survey of 65 DRL developers. The developers confirmed the frequency, severity, and required effort to address identified challenges.
We hope that the reported results in this paper will stimulate the development of DRL quality assurance tools and guide the research community toward solving the identified challenges. 

\section{Data Availability Statement}
The dataset generated during the current study is available in the replication package, which is accessible via \cite{replication_package}.


%
%

\bibliographystyle{spmpsci}      
\bibliography{bibliography}   


\end{document}